\documentclass{article}

\usepackage{docmute}

\usepackage{arxiv}

\usepackage[utf8]{inputenc}
\usepackage[T1]{fontenc}

\usepackage{amsmath}
\usepackage{amssymb}
\usepackage{amsfonts}
\usepackage{amsthm}
\usepackage{bm}
\usepackage{comment}
\usepackage{fancybox}
\usepackage{framed}
\usepackage{color}
\usepackage{graphicx}
\usepackage{cite}
\usepackage{hyperref}
\usepackage{url}

\hypersetup{
    colorlinks = true,
    urlcolor = blue,
    linkcolor = red,
    citecolor = green
}

\usepackage{algorithm}
\usepackage[noend]{algpseudocode}
\usepackage{algorithmicx}

\expandafter\def\expandafter\UrlBreaks\expandafter{\UrlBreaks
  \do\a\do\b\do\c\do\d\do\e\do\f\do\g\do\h\do\i\do\j%
  \do\k\do\l\do\m\do\n\do\o\do\p\do\q\do\r\do\s\do\t%
  \do\u\do\v\do\w\do\x\do\y\do\z\do\A\do\B\do\C\do\D%
  \do\E\do\F\do\G\do\H\do\I\do\J\do\K\do\L\do\M\do\N%
  \do\O\do\P\do\Q\do\R\do\S\do\T\do\U\do\V\do\W\do\X%
  \do\Y\do\Z}

\DeclareMathOperator*{\argmax}{arg\,max}

\algrenewcommand\algorithmicindent{1.0em}

\algnewcommand\algorithmicforeach{\textbf{for each}}
\algdef{S}[FOR]{ForEach}[1]{\algorithmicforeach\ #1\ \algorithmicdo}

\algnewcommand\AlgAnd{\textbf{and} }
\algnewcommand\AlgOr{\textbf{or} }

\algrenewcommand\textproc{}

\algnewcommand{\Initialize}[1]{
	\State \textbf{Initialize:}
 	\State \hspace*{\algorithmicindent}\parbox[t]{0.8\linewidth}{\raggedright #1}}

\title{An FPGA Acceleration and Optimization Techniques for 2D LiDAR SLAM Algorithm}

\author{
    Keisuke Sugiura \\
    Keio University \\
    3-14-1 Hiyoshi, Kohoku-ku, Yokohama, Japan \\
    \texttt{sugiura@arc.ics.keio.ac.jp} \\
    \And
    Hiroki Matsutani \\
    Keio University \\
    3-14-1 Hiyoshi, Kohoku-ku, Yokohama, Japan \\
    \texttt{matutani@arc.ics.keio.ac.jp} \\
}

\hypersetup{
    pdftitle={An FPGA Acceleration and Optimization Techniques for 2D LiDAR SLAM Algorithm},
    pdfsubject={cs.AR},
    pdfauthor={Keisuke Sugiura, Hiroki Matsutani},
    pdfkeywords={SLAM, GMapping, SoC, FPGA}
}

\begin{document}

% Header
\maketitle

% Body

% abst.tex

\begin{abstract}
An efficient hardware implementation for Simultaneous Localization and Mapping (SLAM) methods is of necessity for mobile autonomous robots with limited computational resources.
In this paper, we propose a resource-efficient FPGA implementation for accelerating scan matching computations, which typically cause a major bottleneck in 2D LiDAR SLAM methods.
Scan matching is a process of correcting a robot pose by aligning the latest LiDAR measurements with an occupancy grid map, which encodes the information about the surrounding environment.
We exploit an inherent parallelism in the Rao-Blackwellized Particle Filter (RBPF) based algorithms to perform scan matching computations for multiple particles in parallel.
In the proposed design, several techniques are employed to reduce the resource utilization and to achieve the maximum throughput.
Experimental results using the benchmark datasets show that the scan matching is accelerated by 5.31--8.75$\times$ and the overall throughput is improved by 3.72--5.10$\times$ without seriously degrading the quality of the final outputs.
Furthermore, our proposed IP core requires only 44\% of the total resources available in the TUL Pynq-Z2 FPGA board, thus facilitating the realization of SLAM applications on indoor mobile robots.
\end{abstract}


% intro.tex

\section{Introduction} \label{sec:intro}
Simultaneous localization and mapping (SLAM) technology plays an indispensable role in autonomous robots, such as autonomous driving cars and cleaning robots, and has been a major research topic in robotics over the last two decades.
In order to operate in a previously unknown environment, autonomous robots need to estimate its vehicle pose by matching the sensor observation against the current map, while updating the current map based on the current pose and sensor observation.
Due to this structure of mutual dependence between the robot pose and map, localization and mapping cannot be handled independently from each other.
SLAM algorithms aim to solve these two problems simultaneously.

The Bayes filter-based approach has been widely applied to the SLAM problem.
The variation of Bayes filters including Extended Kalman Filter (EKF) \cite{Dissanayake01} and particle filter are utilized in the process.
FastSLAM \cite{Montemerlo02,Montemerlo03} and GMapping \cite{Grisetti07B} are the most popular methods among particle filter-based approaches and are proven to work well in the literature~\cite{Santos13}.
GMapping is the grid-based LiDAR SLAM based on Rao-Blackwellized Particle Filter (RBPF).
It takes odometry information and measurements from Light Detection and Ranging (LiDAR) sensors as input and generates a sequence of robot poses (trajectory) and an occupancy grid map, which discretize the surrounding environment into equal-sized square cells.

Although SLAM is the key component and basis for autonomous mobile robots, its high computational requirement emerges as a major problem when using SLAM in these robots.
SLAM requires high-end CPUs and sometimes even GPUs to handle massive computations~\cite{Hess16,Nardi15,Gautier19}.
However, there is a situation where these CPUs and GPUs cannot be mounted because of limited power budgets, costs, and physical constraints (size or weight).
Consequently, there exists a strong demand for hardware accelerators to execute SLAM algorithms on such robots.
Hardware offloading brings certain benefits, e.g. performance improvement without additional power consumption.

Particle filter is performed using a set of particles, where each particle carries a single hypothesis of the current state (i.e. robot trajectory and map).
Fortunately, operations on these particles are independent of each other; therefore such an algorithm is suitable for FPGAs with parallel processing capability.
In this paper, an FPGA-based accelerator for GMapping is proposed, by making use of the inherent parallel properties in the algorithm.
Experimental results using benchmark datasets demonstrate that the FPGA accelerator is a feasible solution for improving the throughput without significantly degrading the accuracy.

The rest of this paper is organized as follows.
Section \ref{sec:pre} presents a brief description for GMapping and its theoretical foundation.
In Section \ref{sec:related}, related works for hardware acceleration of RBPF-based SLAM algorithms are reviewed.
In Section \ref{sec:design}, the FPGA accelerator for GMapping is proposed, and its architectural and algorithmic optimizations are described.
Section \ref{sec:impl} illustrates the implementation details.
Evaluation results in terms of throughput, accuracy, resource utilization, and power consumption are shown in Section \ref{sec:eval}.
Section \ref{sec:conc} concludes this paper.

% pre.tex

\section{Preliminaries} \label{sec:pre}
\subsection{Rao-Blackwellized Particle Filter} \label{sec:rbpf}
Rao-Blackwellized Particle Filter (RBPF), an extension of particle filter, is a powerful tool for solving the so-called full SLAM problem~\cite{Grisetti07B}\cite{Thrun05}\cite{Grisetti07A}.
Full SLAM is expressed in the form of the following posterior distribution (\ref{eq:full-slam-problem}) over the state variables consisting of the robot map $m = \{ m_i \}$ and robot trajectory $x_{1 : t} = \{ x_1, \ldots, x_t \}$, conditioned on the sequence of sensor observations $z_{1 : t} = \{ z_1, \ldots, z_t \}$ and robot controls $u_{1 : t} = \{ u_1, \ldots, u_t \}$.
\begin{equation} \label{eq:full-slam-problem}
    p(m, x_{1 : t} | z_{1 : t}, u_{1 : t})
\end{equation}
In particle filters, the above posterior (\ref{eq:full-slam-problem}) is represented by a swarm of particles.
A major drawback is that the number of particles required to sufficiently approximate the posterior grows exponentially with the dimension of the state space.
In the context of SLAM, state variables (i.e. robot pose and map) usually reside in a very high-dimensional space (up to tens of thousands of dimensions).
Therefore, the original particle filter cannot be applied since it would require an enormous amount of particles.
To address this, the posterior (\ref{eq:full-slam-problem}) is decomposed into two terms as shown in Equation (\ref{eq:rbpf-slam}) using the chain rule, which correspond to the trajectory distribution, and the map posterior conditioned on the robot trajectory, respectively \cite{Doucet00}.
\begin{equation} \label{eq:rbpf-slam}
    p(m, x_{1 : t} | z_{1 : t}, u_{1 : t}) = p(x_{1 : t} | z_{1 : t}, u_{1 : t}) p(m | x_{1 : t}, z_{1 : t})
\end{equation}
In RBPF, only the robot trajectory $x_{1 : t}$ is estimated by a particle filter; that is, the set of particles tries to approximate the posterior (\textbf{target}) distribution $P_t$ over the trajectory
\begin{eqnarray} \label{eq:posterior}
    P_t &\equiv& p(x_{1 : t} | z_{1 : t}, u_{1 : t}) \nonumber \\
    &=& \eta \ p(z_t | x_{1 : t}, z_{1 : t - 1}) p(x_t | x_{t - 1}, u_t) P_{t - 1} \nonumber \\
    &\simeq& \eta \ p(z_t | x_t, m) p(x_t | x_{t - 1}, u_t) P_{t - 1}
\end{eqnarray}
The map $m$ is computed deterministically as a function of the trajectory estimate $x_{1 : t}^*$ and the observations $z_{1 : t}$.
In this case, the map distribution $p(m | x_{1 : t}, z_{1 : t})$ can be viewed as a Gaussian with zero variance, where all probability mass is concentrated at the particular point $x_{1 : t} = x_{1 : t}^*$.
Hence, the approximation in Equation (\ref{eq:posterior}) holds, as below:
\begin{eqnarray} \label{eq:approx-map}
    && p(z_t | x_{1 : t}, z_{1 : t - 1}) \nonumber \\
    &=& \int p(z_t | x_t, m') p(m' | x_{1 : t - 1}, z_{1 : t - 1}) dm' \nonumber \\
    &\simeq& p(z_t | x_t, m)
\end{eqnarray}
Each particle individually carries the map as well as trajectory, since the map depends on the estimated trajectory, which differs for each particle.
This factorization yields a significant reduction of the number of particles (i.e. computational cost) because particles are drawn from the relatively low-dimensional space $P_t$ containing robot trajectory only.
The $k$th particle at time $t$, and the particle set at time $t$ are denoted as $Y_t^{[k]} = \{ x_t^{[k]}, m^{[k]}, w_t^{[k]} \}$ and $\mathcal{S}_t = \{ Y_t^{[1]}, \ldots, Y_t^{[M]} \}$ respectively, where $M$ is the number of particles.
RBPF follows the general Sampling Importance Resampling (SIR) algorithm and is outlined by the following four steps.

In the first \textit{sampling} step, a new particle pose $x_t^{[k]}$ is sampled from the Gaussian motion model $p(x_t^{[k]} | x_{t - 1}^{[k]}, u_t)$, which represents the motion uncertainty usually caused by sensor errors, wheel slippages or surface irregularities.
At this point, the set of the particle trajectories $\{ x_{1 : t}^{[k]} \}$ reflects the prior (\textbf{proposal}) distribution $Q_t$ given in Equation (\ref{eq:proposal-rbpf}).
\begin{equation} \label{eq:proposal-rbpf}
    Q_t = p(x_t | x_{t - 1}, u_t) P_{t - 1}
\end{equation}
Then, in \textit{map update} step, the scan data $z_t$ is inserted into each particle map $m^{[k]}$ based on the current particle pose $x_t^{[k]}$, which will be described in detail later.
After that, in \textit{weight update} step, an importance weight associated to each particle $w_t^{[k]}$ is updated based on the ratio between the target $P_t$ and proposal $Q_t$:
\begin{equation} \label{eq:weight-rbpf}
    w_t^{[k]} = w_{t - 1}^{[k]} \frac{P_t}{Q_t}
    \simeq \eta \ w_{t - 1}^{[k]} p(z_t | m^{[k]}, x_t^{[k]})
\end{equation}
In (\ref{eq:weight-rbpf}), $p(z_t | m^{[k]}, x_t^{[k]})$ is the observation likelihood which models the underlying generating process of an observation given the map $m^{[k]}$ and current robot pose $x_t^{[k]}$.
In other words, it represents the consistency of observed data $z_t$ with a map and pose.
Lastly, in \textit{resampling} step, a new generation of particles $\mathcal{S}_t$ is obtained by resampling the particles (allowing duplication) with probability proportional to the importance weights.
Particles with small weights are removed and those with large weights are likely to dominate the entire population.
Particles $\{ x_{1 : t}^{[k]} \}$ now distribute according to the desired posterior distribution $P_t$, which appears in Equation (\ref{eq:rbpf-slam}).
Resampling process is crucial for transforming the particle distribution from prior (proposal) to posterior (target).

\subsection{GMapping} \label{sec:gmapping}
GMapping is classified as the RBPF-SLAM algorithm and is commonly used among the robotics community.
It periodically retrieves the latest robot control $u_t$ and scan data $z_t = \{ z_t^i \}$ captured from a LiDAR sensor.
It then builds a planar occupancy grid map $m$, in which each grid cell contains a probability that the cell is occupied by an object.
A single observation $z_t^i = [r_t^i, \theta_t^i]^\top$ is comprised of distance $r_t^i$ and angle $\theta_t^i$ with respect to the sensor.

GMapping employs two strategies to reduce the computational burden: improved proposal and adaptive resampling.
In the \textit{sampling} step, a new particle pose $x_t^{[k]}$ is drawn from the altered distribution (\ref{eq:proposal-distribution}) instead of the raw odometry motion model $p(x_t | x_{t - 1}, u_t)$.
\begin{equation} \label{eq:proposal-distribution}
    p(x_t | m, x_{t - 1}, z_t, u_t) = \frac{p(z_t | m, x_t) p(x_t | x_{t - 1}, u_t)}{\displaystyle \int p(z_t | m, x) p(x | x_{t - 1}, u_t) dx}
\end{equation}
The above distribution (\ref{eq:proposal-distribution}) also takes into account the latest observation $z_t$ and is more peaked than the ordinary motion model, thereby providing a highly accurate pose $x_t$ \cite{Grisetti07B}.
To perform a sampling based on Equation (\ref{eq:proposal-distribution}), the robot pose $x_t'^{[k]}$ is initially sampled from the motion model $p(x_t | x_{t - 1}, u_t)$ and then is refined so that the current scan $z_t$ and map $m^{[k]}$ maximally overlap each other.
This alignment is called \textit{scan matching}, and involves the maximization of the likelihood function formalized as below.
\begin{equation} \label{eq:scan-matching}
    x_t^{[k]} = \argmax_x p(x | m^{[k]}, z_t, x_t'^{[k]})
\end{equation}
It leads the particles to be located in a more meaningful area with higher observation likelihood, thus reducing the number of particles and improving algorithmic efficiency.
The proposal now takes the following form
\begin{equation} \label{eq:proposal-gmapping}
    Q_t' = p(x_t | m, x_{t - 1}, z_t, u_t) P_{t - 1}.
\end{equation}
The importance weight is then computed as follows
\begin{eqnarray} \label{eq:importance-weight}
    w_t^{[k]} &=& w_{t - 1}^{[k]} \frac{P_t}{Q_t'} \nonumber \\
    &=& \eta \ w_{t - 1}^{[k]} \frac{p(z_t | x_t^{[k]}, m^{[k]}) p(x_t^{[k]} | x_{t - 1}^{[k]}, u_{t - 1})}{p(x_t^{[k]} | m^{[k]}, x_{t - 1}^{[k]}, z_t, u_t)} \nonumber \\
    &=& \eta \ w_{t - 1}^{[k]} \int p(z_t | m^{[k]}, x) p(x | x_{t - 1}^{[k]}, u_t) dx.
\end{eqnarray}
Since the observation likelihood has a much smaller variance than the motion model, the integral above may be evaluated around the maximum of the likelihood, $x_t^{[k]}$, which is already obtained as a result of scan matching.
Consequently, the weight computation (\ref{eq:importance-weight}) is further simplified to Equation (\ref{eq:importance-weight-approx}).
\begin{equation} \label{eq:importance-weight-approx}
    w_t^{[k]} \simeq \eta \ w_{t - 1}^{[k]} p(z_t | m^{[k]}, x_t^{[k]})
\end{equation}
Resampling is only performed when the effective sample size in Equation (\ref{eq:effective-sample-size}) falls below the threshold value $M_\mathrm{th}$.
\begin{equation} \label{eq:effective-sample-size}
    M_\mathrm{eff} = \frac{1}{\sum_k \left( w_t^{[k]} \right)^2}
\end{equation}
$M_\mathrm{eff}$ can be interpreted as the accuracy of the proposal.
It reaches its maximum value $M$ when all weights are identical ($w_t^{[k]} = M^{-1}$), that is, the proposal distribution fully reflects the target distribution.
An excessive variance of the importance weights incurs a small $M_\mathrm{eff}$.
Especially when $M_\mathrm{eff}$ is large, resampling is unnecessary since the current particle set is assumed to represent the target distribution effectively.
The adaptive resampling technique enables to retain the diversity of hypotheses and thus mitigates the risk of the particles around the correct state being removed, also known as particle deprivation (depletion).

Algorithm \ref{alg:gmapping} summarizes the overall algorithm of GMapping, where the symbol $\oplus$ denotes the composition operator~\cite{Kuemmerle09} and $\varepsilon$ is the zero-mean Gaussian noise.
\begin{algorithm}[h]
    \caption{GMapping Algorithm}
    \label{alg:gmapping}
    \begin{algorithmic}[1]
        \Function{GMapping}{\null}
            \State $t \gets 1, \quad \mathcal{S}_0 \gets \varnothing$
            \For{$k = 1, \ldots, M$} \Comment{Initialize particle set}
                \State $\mathcal{S}_0 \gets \mathcal{S}_0 \cup \left\{ x_0, m_0, M^{-1} \right\}$
                \Comment{Set initial pose, empty grid map, and initial weight}
            \EndFor
            \While{$\left\{ u_t, z_t \right\}$ exists} \Comment{Consume sensor data}
                \State $\mathcal{S}_t \gets $ \Call{Process}{$\mathcal{S}_{t - 1}, z_t, u_t$}, $\quad t \gets t + 1$
            \EndWhile
            \State $k^* \gets \argmax_k \left\{ w_t^{[k]} \right\}$
            \Comment{Choose the best particle with largest importance}
            \State \Return $x_{1 : t}^{[k^*]}, m^{[k^*]}$
            \Comment{Return the most plausible trajectory and map}
        \EndFunction \newline

        \Function{Process}{$\mathcal{S}_{t - 1}, z_t, u_t$}
            \State $\mathcal{S}_t = \varnothing$ \Comment{Initialize new particle set}
            
            \ForEach{$Y_{t - 1}^{[k]} \in \mathcal{S}_{t - 1}$}
                \State $x' \gets x_{t - 1}^{[k]} \oplus u_t + \varepsilon$ \Comment{Initial guess}
                \State $x_t^{[k]} \gets \argmax_{x} p(x | m^{[k]}, z_t, x')$ \Comment{Scan matching}
                \State $m^{[k]} \gets $ \Call{AddScan}{$m^{[k]}, x_t^{[k]}, z_t$} \Comment{Update map}
                \State $w_t^{[k]} \gets \eta \ w_{t - 1}^{[k]} \displaystyle \int p(x | x_{t - 1}^{[k]}, u_t) p(z_t | x, m^{[k]}) dx$
                \Comment{Update weight}
                \State $\mathcal{S}_t \gets \mathcal{S}_t \cup \left\{ x_{1 : t}^{[k]}, m^{[k]}, w_t^{[k]} \right\}$ \Comment{Add to new particle set}
            \EndFor \medskip

            \State $M_\mathrm{eff} = \left[ \sum_k \left( w_t^{[k]} \right)^2 \right]^{-1}$
            \Comment{Compute effective sample size}
            \If{$M_\mathrm{eff} < M_\mathrm{th}$}
                \State $\mathcal{S}_t \gets $ \Call{Resample}{$\mathcal{S}_t$} \Comment{Resample if necessary}
            \EndIf \medskip
            \State \Return $\mathcal{S}_t$
        \EndFunction
    \end{algorithmic}
\end{algorithm}

The function \textrm{AddScan}($m, x_t, z_t$) incorporates the scan data $z_t$ into the map $m$ using the robot position $x_t$.
It transforms each scan $z_t^i = \left[ r_t^i, \theta_t^i \right]^\top$ from the sensor coordinate to the map coordinate and computes the hit point (also referred to as the beam endpoint) $p_t^i$.
Then it determines the \textit{hit} grid cell that contains $p_t^i$ and \textit{missed} grid cells that lie on the straight line connecting $p_t^i$ and $x_t$ using Bresenham's algorithm.
Binary Bayes filter is applied to these grid cells and their occupancy probabilities are incrementally updated.
The probability values associated with \textit{missed} cells are lowered since they are less likely to be obstructed (laser rays just went through these cells), and opposite for the \textit{hit} cell.

% related.tex

\section{Related Work} \label{sec:related}
There are several works on accelerating RBPF-based SLAM methods for embedded platforms by exploiting their parallel nature~\cite{Par11,Llofriu13,Gouveia15,Abouzahir16,Sileshi16}.
Abouzahir \textit{et al.} quantitatively analyzed execution times of SLAM algorithms under varying parameter settings and concluded that FastSLAM 2.0 is preferable for the low-cost embedded systems in terms of the real-time performance and consistency of output results~\cite{Abouzahir18}.
Their implementation of the Monocular FastSLAM 2.0 targeting CPU-FPGA heterogeneous architectures outperformed those run on high-end CPU or GPU and demonstrated the feasibility of FPGA as an accelerator in the domain of SLAM.
FastSLAM 2.0 is also a variant of the RBPF-based method as GMapping~\cite{Montemerlo03}.
The primary difference is that FastSLAM 2.0 builds a feature-based map, consisting of the features of landmarks recognized by robots, while GMapping constructs a grid-based map.

Both map representations are widely used; however, the former requires the feature extraction and detection from sensor inputs, i.e., prior knowledge about the environment structure.
The main advantage of the grid-based map is its flexibility, meaning that it can represent arbitrary objects and thus no assumption of the environment is needed~\cite{Wurm10}.
Also, occupancy state at any location is easily obtainable owing to its dense data structure, making it a convenient format for other tasks such as navigation and motion planning, which are based on pathfinding algorithms.
From the aspect of the scan matching using LiDAR data, the matching between a scan and a grid map (often referred to as \textit{scan-to-map}) generally produces accurate and robust alignments than the matching between two scans (\textit{scan-to-scan})~\cite{Hess16}.

The major drawback of the grid-based map is that it demands a large amount of memory in exchange for its dense representation~\cite{Wurm10}.
This problem is even more critical in RBPF-based SLAM, because each particle keeps its individual map, which means that the memory consumption grows quadratically to the map size and also proportionally to the number of particles.
However, several techniques are proposed to mitigate the problem by sharing a part of grid map among multiple particles~\cite{Grisetti07A}\cite{Schroeter07}\cite{Jo18}, exploiting the implicit redundancy in the particle maps.
That is, multiple identical copies of a single particle map are created during a \textit{resampling} process, but only tiny fractions of them are modified and large parts remain unchanged.
In our software implementation, the map sharing technique similar to \cite{Schroeter07} is applied to reduce the memory consumption and to increase the maximum number of particles.
These techniques are effective especially when the RBPF-SLAM is being run on a resource-limited platform.
From the above considerations, the grid-based approach is focused in this paper.
To the best of our knowledge, this is the first work that presents FPGA design for grid-based RBPF-SLAM.

Gouveia \textit{et al.} proposed a multithreaded version of GMapping using the OpenMP library, and high map precision was gained by increasing the number of particles without sacrificing the latency~\cite{Gouveia14}.
Li \textit{et al.} also examined an acceleration of GMapping leveraging several parallel processing libraries~\cite{QiuchengLi18}.
The above-mentioned works focus on GMapping acceleration from the software aspects.
In this paper, on the other hand, we investigate the FPGA implementation of GMapping for the first time and propose optimization methods to achieve resource efficiency and high-performance.

% For instance, a well-known \textit{scan-to-scan} method called Iterative Closest Points (ICP) \cite{Besl92} involves the data association process, where every scan point is associated with the closest scan point in the reference scan.
% It is susceptible to outliers or false data associations~\cite{Olson15}.
% As described later, the \textit{scan-to-map} matching does not require such association; it simply projects the scan data onto the map.

% design.tex

\section{Design Optimization} \label{sec:design}
In Section \ref{sec:parallel-processing}, we first provide a reason for choosing the scan matching part as a target of the hardware acceleration.
We then thoroughly describe the algorithm for scan matching in Section \ref{sec:greedy-endpoint-matching} and three optimization techniques adopted in the hardware implementation in Sections \ref{sec:map-compression}-\ref{sec:simplified-score-calculation}.

\subsection{Parallelization of Scan Matching} \label{sec:parallel-processing}
\begin{figure}[h]
    \centering
    \includegraphics[width=0.5\linewidth]{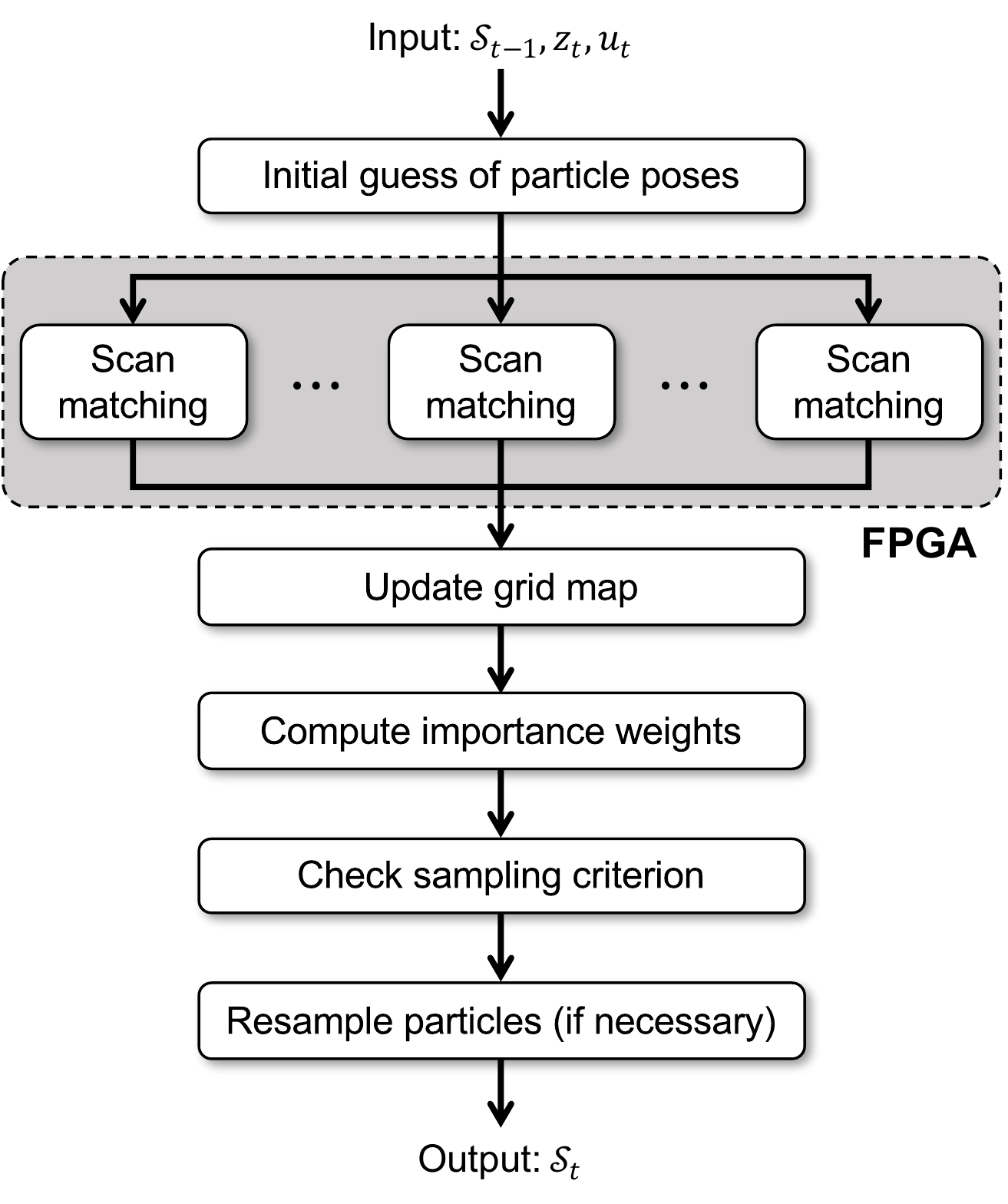}
    \caption{Parallelization of Scan Matching}
    \label{fig:parallel-processing}
\end{figure}

% This section provides the detailed description of the FPGA accelerator for scan matching computations.
As described in Section \ref{sec:gmapping}, the algorithm is divided into five main parts: \textit{initial guess}, \textit{scan matching}, \textit{map udpate}, \textit{weight update}, and \textit{resampling}.
Its notable feature is that all the operations except resampling can be performed simultaneously for multiple particles.
Scan matching is the process of superimposing a scan on a grid map, i.e. it tries to find the most suitable alignment so that a map and a scan projected onto a map maximally overlap each other.
It inevitably becomes time-consuming and computationally intensive~\cite{Olson15}, since a large number of calculations (especially coordinate transformations) and random accesses to the map are required.
Performance evaluations in Section \ref{sec:eval} reveal that scan matching accounts for up to 90 \% of the total execution time, clearly posing a major bottleneck.
Scan matching is the most reasonable candidate for hardware acceleration in terms of the expected performance gain.
In this paper, as illustrated in Figure \ref{fig:parallel-processing}, scan matching is executed in parallel on an FPGA device and other necessary computations are handled on the CPU side, thus utilizing the heterogeneous SoC architecture.

\subsection{Greedy Endpoint Matching Algorithm} \label{sec:greedy-endpoint-matching}
The software implementation used in this paper is based on the open-source package provided by OpenSLAM~\cite{OpenSLAM}.
In the OpenSLAM GMapping package, a metaheuristic hill-climbing based algorithm called Greedy Endpoint Matching~\cite{Montemerlo03B} is executed during the scan matching process.
It is worth noting that more sophisticated algorithms like branch-and-bound based method~\cite{Hess16} and correlation-based method~\cite{Olson09} can be applied for scan matching.
Although the hill-climbing method has a weakness that its performance is negatively affected by the poor initial estimates and is susceptible to local optima~\cite{Olson09}, a comparison of the scan matching algorithms' performance is outside the scope of this paper.

The hill-climbing algorithm corrects a particle pose $x_t$ by aligning a scan data $z_t = \left\{ z_t^i \right\}$ with a map $m$.
More concretely, a particle pose $x_t$ that maximizes a matching score $s(x_t, m, z_t)$ is continually explored until a convergence is reached.
The matching score is regarded as the observation likelihood $p(x_t | m, z_t, x_t')$ as mentioned in Section \ref{sec:pre}, where $x_t'$ denotes an initial estimate of a particle pose.
In each iteration, the algorithm chooses an axial direction that most improves the score, and then the particle pose is updated by a small step along that direction.
The update step, which is analogous to a learning rate in gradient descent optimization, is halved if the score is not improved and no feasible direction is found, and the algorithm ends if a convergence criterion is met (i.e. the update step becomes sufficiently small).
The score $s(x_t, m, z_t)$ is calculated according to the following equation.
\begin{equation} \label{eq:scan-matching-score}
    s(x_t, m, z_t) = \sum_i \exp \left\{ -\frac{\left( d_t^i \right)^2}{2 \sigma^2} \right\} = \sum_i u(d_t^i),
\end{equation}
where $\sigma$ is the predefined standard deviation and summand $u(d_t^i)$ is the score for $i$th measurement $z_t^i$.
The $d_t^i$ denotes the distance between the $i$th scan point $p_t^i$ (described in Section \ref{sec:pre}) and its closest obstacle registered in the map $m$.
A smaller value of $d$ implies a small misalignment between the observation $z_t$ and the map $m$.
Scan point $p_t^i$ of the $i$th observation $z_t^i = [r_t^i, \theta_t^i]^\top$ is computed by the coordinate transformation from the sensor frame to the map frame under the current pose $x_t = [\xi_t^x, \xi_t^y, \xi_t^\theta]^\top$ as follows.
\begin{equation} \label{eq:scan-to-map}
    p_t^i = \left[ \begin{array}{l}
        \xi_t^x + r_t^i \cdot \cos(\xi_t^\theta + \theta_t^i) \\
        \xi_t^y + r_t^i \cdot \sin(\xi_t^\theta + \theta_t^i)
    \end{array} \right] \in \mathbb{R}^2
\end{equation}

\begin{figure}[h]
    \centering
    \includegraphics[width=0.6\linewidth]{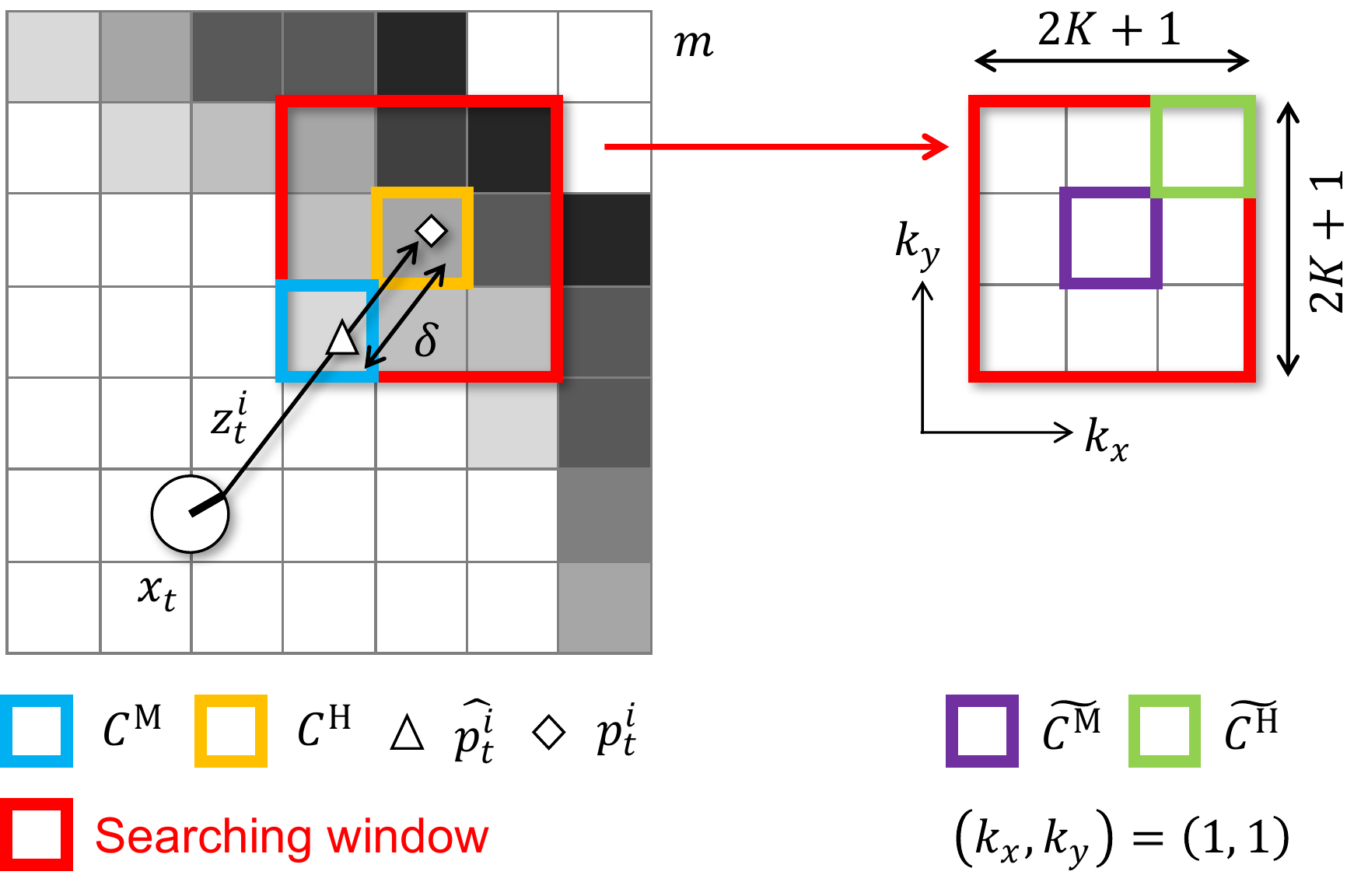}
    \caption{Scan Point and its Surroundings}
    \label{fig:hit-and-missed-cell}
\end{figure}

The naive yet stable algorithm to find the minimum distance $d_t^i$ is summarized in Algorithm \ref{alg:minimum-distance}.
$\gamma(x^\mathrm{m}): \mathbb{R}^2 \to \mathbb{Z}^2$ is a function that converts the position in the map frame $x^\mathrm{m} = \left[ \xi_x^\mathrm{m}, \xi_y^\mathrm{m} \right]^\top$ to the corresponding grid cell index.
It is formulated as
\begin{equation} \label{eq:map-to-cell}
    \gamma(x^\mathrm{m}) = \left[ \begin{array}{l}
        \lfloor (\xi_x^\mathrm{m} - o_x) / \Delta \rfloor \\
        \lfloor (\xi_y^\mathrm{m} - o_y) / \Delta \rfloor
    \end{array} \right] \in \mathbb{Z}^2,
\end{equation}
where $\Delta$ is a map resolution (grid cell size) and $[o_x, o_y]^\top$ denotes the position of the map origin (the position that corresponds to the grid cell with a minimum index $(0, 0)$), respectively.
$\gamma^{-1}(C_x, C_y)$ is the inverse of $\gamma$, written as
\begin{equation} \label{eq:cell-to-map}
    \gamma^{-1}(C_x, C_y) = \left[ \begin{array}{l}
        o_x + C_x \Delta \\
        o_y + C_y \Delta
    \end{array} \right] \in \mathbb{R}^2.
\end{equation}
Algorithm \ref{alg:minimum-distance} first calculates the scan point $p_t^i$ and its closest grid cell $C^\mathrm{H}$ for each scan $z_t^i$.
It then calculates $\widehat{p}_t^i$ and $C^\mathrm{M}$ in the same way.
$\widehat{p}_t^i$ is the point that is closer to the sensor by $\delta$ than the scan point $p_t^i$.
The cell $C^\mathrm{M}$ is therefore presumed to be unoccupied and missed by the beam (i.e. $C^\mathrm{M}$ should belong to the set of \textit{missed} grid cells, because the laser beam passed through the cell $C^\mathrm{M}$).
Figure \ref{fig:hit-and-missed-cell} (left) shows an example of the positional relationship between $p_t^i$ and $\widetilde{p}_t^i$.

After that, it attempts to establish the matching between the observation $z_t^i$ and the map $m$.
It utilizes a square searching window of $(2K + 1) \times (2K + 1)$ cells, centered at the $C^\mathrm{H}$ (see Figure \ref{fig:hit-and-missed-cell} (left)).
In our implementation, the radius $K$ is currently set to $1$, yielding the $3 \times 3$ square searching window.
Every grid cell covered by the window is considered a candidate for containing the beam endpoint $p_t^i$.
That is, $p_t^i$ might not reside in the $C^\mathrm{H}$ but in proximity to the $C^\mathrm{H}$ because of the accumulated error in $x_t$ or the perturbation in measurement $z_t^i$.
The searching window is to allow these errors and to consider the case where $p_t^i$ does not exactly correspond to $C^\mathrm{H}$.

Each cell in the searching window and its associated occupancy probability are denoted as $\widetilde{C}^\mathrm{H}$ and $p^\mathrm{H}$, respectively.
The index of $\widetilde{C}^\mathrm{H}$ is given by adding a relative offset $(k_x, k_y)$ to $C^\mathrm{H}$ (refer to Figure \ref{fig:hit-and-missed-cell} (right)).
The same applies to $p^\mathrm{M}$ and $\widetilde{C}^\mathrm{M}$.
For each cell $\widetilde{C}^\mathrm{H}$, it is tested whether two values $p^\mathrm{H}$ and $p^\mathrm{M}$ are within the desired ranges: $(T, 1]$ and $[0, T)$.
These criteria are derived from the fact that $\widetilde{C}^\mathrm{H}$ and $\widetilde{C}^\mathrm{M}$ should be \textit{hit} and \textit{missed} cell.
If $\widetilde{C}^\mathrm{H}$ satisfies these criteria, $C^\mathrm{H}$ is the appropriate matching candidate and is expected to accommodate the scan point $p_t^i$, meaning that $p_t^i$ actually resides in $\widetilde{C}^\mathrm{H}$ and not $C^\mathrm{H}$.
The distance between $p_t^i$ and $p'$ is then calculated, where $p_t^i$ is the scan point obtained from the current pose $x_t$ using Equation (\ref{eq:scan-to-map}), and $p'$ is its corresponding point found on the map $m$, respectively.
The minimum distance is selected for $d_t^i$ if multiple grid cells satisfy the criteria.
Checking the value of $p^\mathrm{M}$, which is expected to be lower than the $p^\mathrm{H}$, effectively avoids the false matching and hence contributes to the robustness.

\begin{algorithm}[h]
    \caption{Calculation of $d_t^i$}
    \label{alg:minimum-distance}
    \begin{algorithmic}[1]
        \Function{FindMinimumDistance}{$x_t, m, z_t^i$}
            \State $p_t^i \gets \left[ \begin{array}{l}
            \xi_t^x + r_t^i \cos(\xi_t^\theta + \theta_t^i) \\
            \xi_t^y + r_t^i \sin(\xi_t^\theta + \theta_t^i)
            \end{array} \right]$
            \Comment{Compute scan point}
            \State $\widehat{p}_t^i \gets \left[ \begin{array}{l}
            \xi_t^x + (r_t^i - \delta) \cos(\xi_t^\theta + \theta_t^i) \\
            \xi_t^y + (r_t^i - \delta) \sin(\xi_t^\theta + \theta_t^i)
            \end{array} \right]$
            \Comment{Compute point that seems unoccupied}
            \State $C^\mathrm{H} \gets \gamma(p_t^i)$ \Comment{Compute \textit{hit} cell index}
            \State $C^\mathrm{M} \gets \gamma(\widehat{p}_t^i)$ \Comment{Compute \textit{missed} cell index}
            \State $d^* \gets \infty$ \Comment{Initialize minimum distance} \newline
            \For{$k_x = -K, \ldots, K$}
                \For{$k_y = -K, \ldots, K$}
                \Comment{For each cell in searching window}
                    \State $\widetilde{C}^\mathrm{H} \gets (C_x^\mathrm{H} + k_x, C_y^\mathrm{H} + k_y), \ p^\mathrm{H} \gets m(\widetilde{C}^\mathrm{H})$
                    \State $\widetilde{C}^\mathrm{M} \gets (C_x^\mathrm{M} + k_x, C_y^\mathrm{M} + k_y), \ p^\mathrm{M} \gets m(\widetilde{C}^\mathrm{M})$
                    \Comment{Check occupancy probabilities of candidate cells}
                    \If{$p^\mathrm{H} > T$ \AlgAnd $p^\mathrm{M} < T$} \label{alg:minimum-distance-comparison}
                        \State $p' \gets \gamma^{-1}(\widetilde{C}^\mathrm{H}), \ d^* \gets \min(d^*, \left| p_t^i - p' \right|)$
                        \Comment{Update minimum distance if criteria met}
                    \EndIf
                \EndFor
            \EndFor \newline
            \State \Return $d^*$
        \EndFunction
    \end{algorithmic}
\end{algorithm}

\begin{figure}[h]
    \centering
    \includegraphics[width=0.7\linewidth]{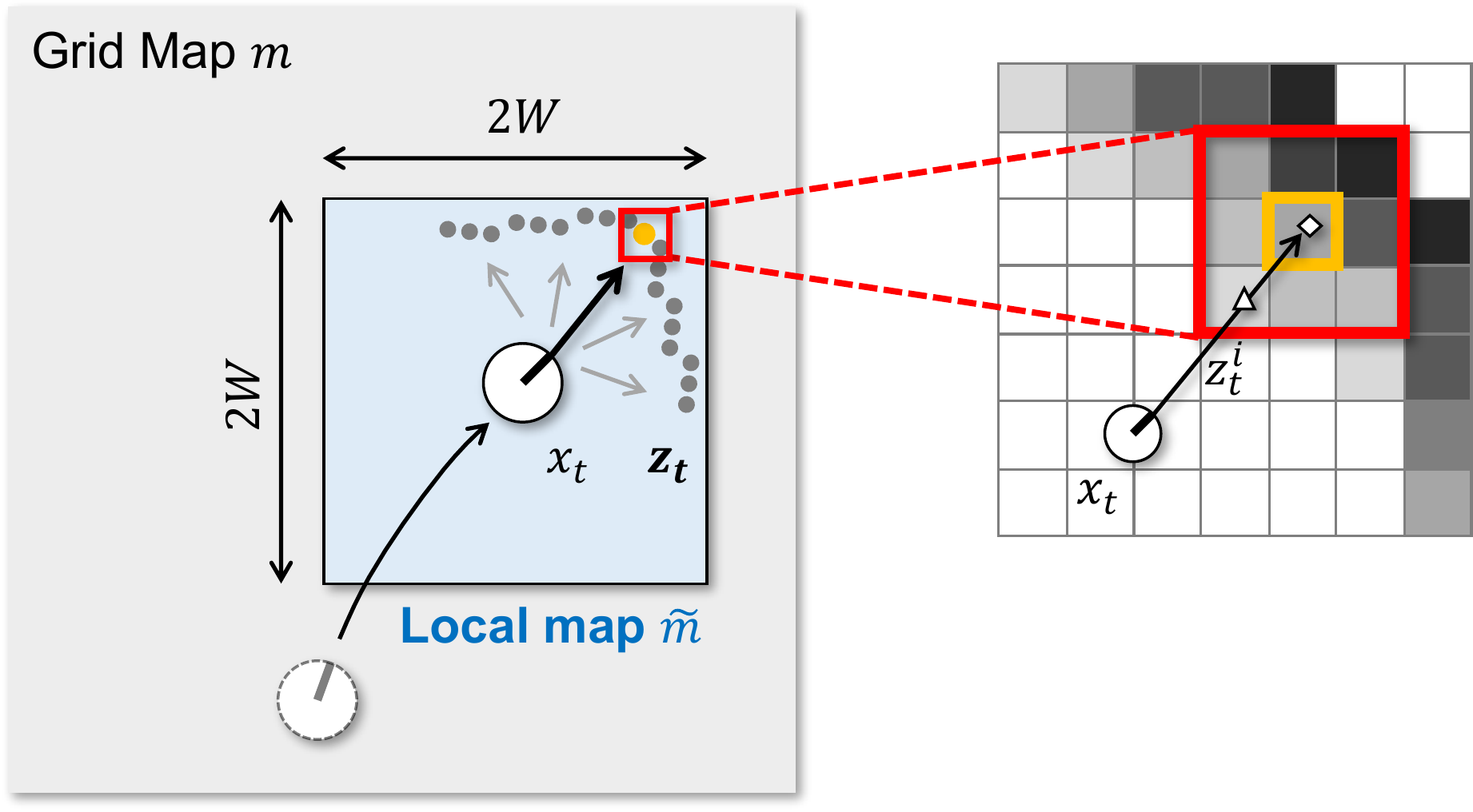}
    \caption{Entire Grid Map and Local Map}
    \label{fig:submap-overview}
\end{figure}

The optimizations to realize the resource-efficient implementation are threefold: (a) map compression, (b) efficient access to map data, and (c) simplified score calculation.

\subsection{Map Compression} \label{sec:map-compression}
The map resolution $\Delta$ is preferred to be set to a smaller value, e.g. 0.01 \textrm{m} or 0.05 \textrm{m}, since it directly affects the accuracy of the output map.
More importantly, the RBPF-based approach requires map hypotheses to be maintained individually on each particle.
The amount of memory needed to store the map increases approximately to the square of the map size, inversely to the square of the map resolution $\Delta$, and also proportional to the number of particles $M$.
Typically, it ranges in the order of hundreds of megabytes, especially when a considerable number of particles are used to deal with a mapping in a relatively large environment.
On an FPGA platform with limited hardware resources, the amount of FPGA on-chip memory (BRAM) is not enough for even storing one single map, and thus frequent data transfer between the BRAM and an on-board DRAM will be required.
In addition, transferring such amount of data imposes a massive overhead, which potentially outweighs the advantage of hardware acceleration.
As a result, an effective way of reducing the map size should be devised.

Considering the physical principle of a LiDAR sensor, it is immediately apparent that only a fraction of the mapped area is observable from a sensor at any iteration.
This indicates that the \textit{local} map covering only the surrounding of the robot can be utilized during the scan matching process instead of the entire map, a significant part of which is eventually not used.
Local map $\widetilde{m}$ is essentially a cropped version of the original map $m$.
Local map for $k$th particle $\widetilde{m}^{[k]}$ is constructed by clipping an area of the predetermined size of $2W \times 2W$ grid cells from the map $m^{[k]}$, centering on the grid cell $(C_x, C_y)$ corresponding to the current pose $x_t^{[k]}$ (see Figure \ref{fig:submap-overview}).
\begin{equation} \label{eq:local-map-definition}
    \widetilde{m}^{[k]} = \left\{ m^{[k]}(C_x + k_x, C_y + k_y) | k_x, k_y \in \left[ -W, W \right) \right\}
\end{equation}
This amounts to the approximation of proposal distribution $p(x_t | m, x_{t - 1}, z_t, u_t)$ by substituting the map $m$ with the local map $\widetilde{m}$~\cite{Grisetti07A}.
In the current implementation, $\Delta$ and $W$ are set to 0.05 \textrm{m} and 128, respectively, making a local map 12.8 \textrm{m} square.
$W$ should be selected so that almost every scan point fits inside the local map; otherwise, the accuracy of scan matching is seriously lost.
The scan points that are out of the local map are not taken into account in the score evaluation (Equation (\ref{eq:scan-matching-score})) and the algorithm greatly suffers from the resulting inaccurate score.
In an environment densely occupied with obstacles, smaller $W$ is applicable, since the distance to the nearest obstacle (obtained as a scan data from a laser scanner) tends to become relatively shorter.
Use of local maps clearly reduces both hardware amount and data transfer latency.
As a side benefit, each map can be viewed as a fixed-size 2D array from the FPGA side, thus facilitating data retrieval and processing.
On the software, the map is implemented as a variable-sized array and is dynamically expanded when a robot enters previously unexplored areas, whereas the size of the local map remains unchanged.

\begin{figure}[h]
    \centering
    \includegraphics[width=0.5\linewidth]{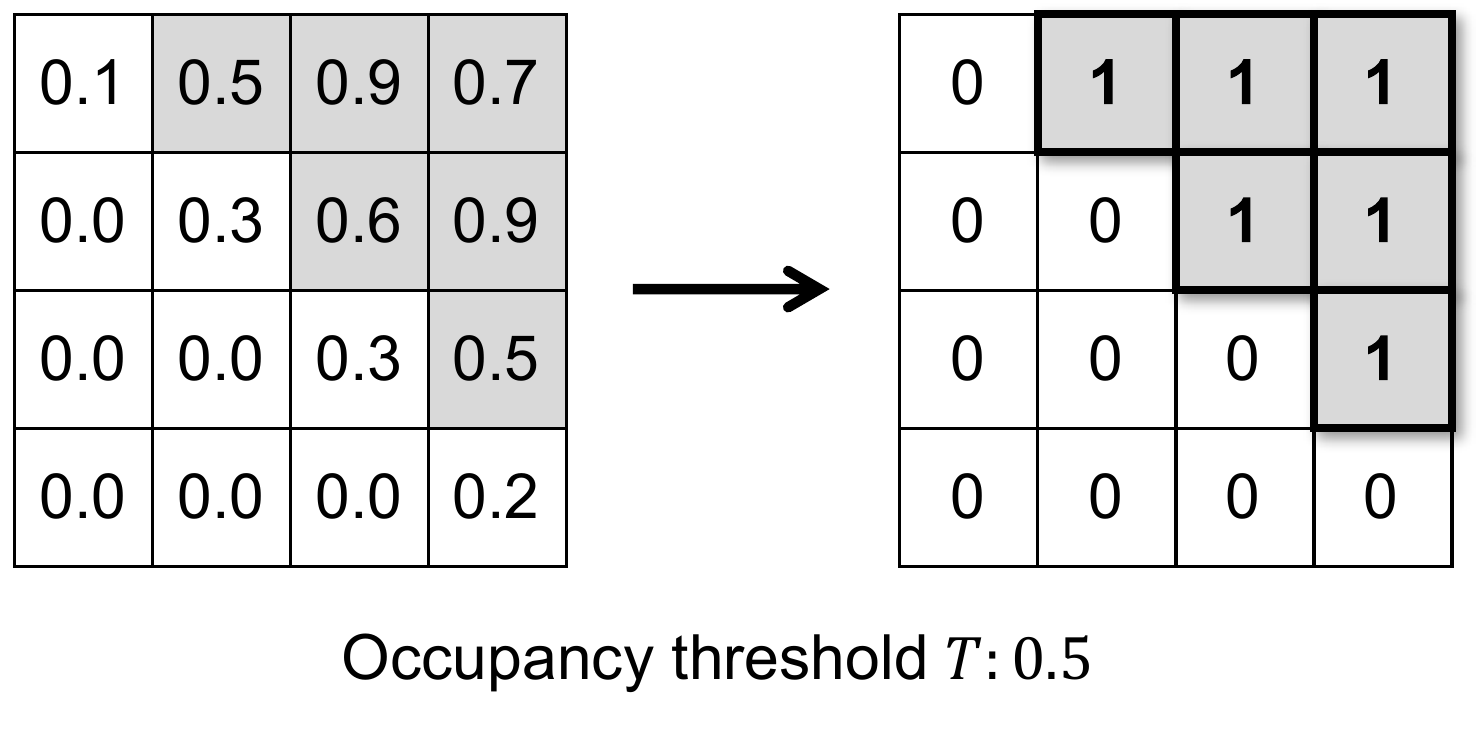}
    \caption{Local Map Binarization}
    \label{fig:submap-binarization}
\end{figure}

An occupancy value is stored in a double-precision floating-point format in the software implementation.
According to Algorithm \ref{alg:minimum-distance}, however, one can find that the floating-point representation is completely redundant since the value is only used for the comparison against the occupancy threshold $T$; the value itself is not of interest.
For this reason, occupancy values can be quantized into 1-bit values by performing this comparison before being fed to the FPGA scan matcher core (Figure \ref{fig:submap-binarization}).
This binarization reduces resource usage by up to 64$\times$ with no accuracy loss and it finally becomes feasible to store local maps for multiple particles on BRAM blocks for parallel processing.
Also, time-consuming DRAM accesses from inside of an FPGA are fully eliminated and the data transfer overhead is substantially reduced.
Overall latency is also reduced in the way that a comparison between two floating-point numbers (appears in Line \ref{alg:minimum-distance-comparison} in Algorithm \ref{alg:minimum-distance}) is turned into a simple bit operation.

\subsection{Efficient Access to Map Data} \label{sec:efficient-access-to-map-data}
As mentioned in Section \ref{sec:greedy-endpoint-matching}, the searching window has the size of $3 \times 3$ grid cells.
Under this setting, one can observe that the single execution of Algorithm \ref{alg:minimum-distance} results in eighteen consecutive accesses to the grid cells in map $m$; nine for the \textit{hit} cells $\widetilde{C}^\mathrm{H}$ and the other nine for the \textit{missed} cells $\widetilde{C}^\mathrm{M}$.
Minimizing the latency for the map data acquisitions (i.e. BRAM accesses) is crucial because it resides in the innermost part of the scan matching algorithm and thus it directly affects the entire performance of the IP core.

\begin{figure}[h]
    \centering
    \includegraphics[width=0.8\linewidth]{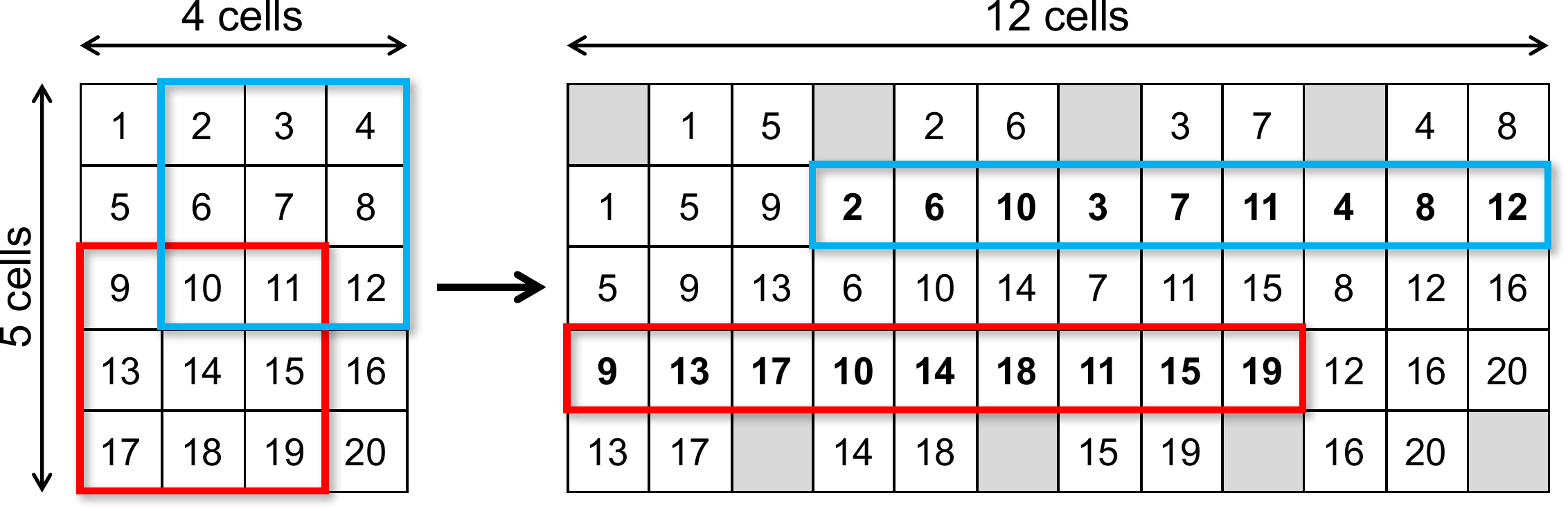}
    \caption{Layout of Map Data on BRAM}
    \label{fig:submap-layout}
\end{figure}

An example of the typical access pattern that occurs when sweeping a single searching window (consisting of nine elements) is shown in Figure \ref{fig:submap-layout} (left).
In this case, in order to obtain all nine elements in a single cycle, the map data (2D array) needs to be completely partitioned along both dimensions, thereby eating up valuable memory resources.
In our FPGA scan matcher, however, the algorithm does not follow the above access pattern; instead, it accesses the data along a horizontal direction (with the vertical position being fixed) as depicted in Figure \ref{fig:submap-layout} (right).
Apparently, the amount of memory to store the map increases by $3 \times$, since the map now needs to contain duplicate elements to achieve this dedicated access pattern.
The primary advantage of the modification of the data layout is that the algorithm can query a searching window within a single clock cycle by partitioning the map along a horizontal axis only, without the necessity of full partitioning.
Avoiding the unnecessary partitioning is effective for reducing the resource usage.
Despite of the $3 \times$ increase of the memory footprint caused by allowing redundancy, it is still possible to keep multiple grid maps on BRAMs by combining the map binarization and cropping technique presented in Section \ref{sec:map-compression}.
In this way, the minimum latency for the map data accesses is achieved, mitigating the negative effects on the resource utilization.

\subsection{Simplified Score Calculation} \label{sec:simplified-score-calculation}
% Under the independence assumption amongst measurements $z_t = \{ z_t^i \}$ as shown in Equation (\ref{eq:measurement-independence}), the scan matching score is calculated separately for each measurement $z_t^i$ as Equation (\ref{eq:scan-matching-score}), depending on its closeness to the obstacle $d(x, z_t^i, m)$.
% \begin{equation} \label{eq:measurement-independence}
%     p(z_t | m, x_t) = \prod_i p(z_t^i | m, x_t)
% \end{equation}
According to Algorithm \ref{alg:minimum-distance}, $d'$ is essentially the distance $d' = |p_t^i - p'|$ between the two grid cells $C^\mathrm{H}$ and $\widetilde{C}^\mathrm{H}$, which correspond to the scan point $p_t^i$ and its actual point $p'$ on the map $m$ as described in Section \ref{sec:map-compression}, respectively.
Inspecting the following Equation (\ref{eq:approx-dist}) reveals that $d$ can be computed from only the offsets $k_x, k_y$, and map resolution $\Delta$ by approximating $p_t^i$ with $\gamma^{-1}(C^\mathrm{H})$; hence the absolute positions $p_t^i, p'$ are unneeded.
\begin{eqnarray} \label{eq:approx-dist}
    d' &=& \left| p_t^i - p' \right| \nonumber \\
    &\simeq& \left| \gamma^{-1}(C^H) - \gamma^{-1}(\widetilde{C}^H) \right| \nonumber \\
    &=& \sqrt{(C_x^H - \widetilde{C}_x^H)^2 + (C_y^H - \widetilde{C}_y^H)^2} \Delta \nonumber \\
    &=& \sqrt{k_x^2 + k_y^2} \Delta
\end{eqnarray}
It turns out that $d'$ and $u(d') = \exp(-d'^2 / 2 \sigma^2)$ are discrete functions of relative offsets $k_x, k_y \in [-K, K]$.
Note that $u(d')$ is a scan matching score for a single observation that appears in Equation (\ref{eq:scan-matching-score}).
A lookup table of size $(2K + 1)^2$ that contains the Gaussian $u(d')$ of every possible distance $d'$ (i.e. every possible combinations of offsets $k_x, k_y$) can be computed beforehand.
This lookup table can be fully partitioned and mapped as registers, since it consists of only nine elements when $K = 1$.
This precomputation enables the effective evaluation of the score $s(x, m, z_t)$ since the computation of the Gaussian function in Equation (\ref{eq:scan-matching-score}) is replaced by the single query to the lookup table entry.

% impl.tex

\section{Implementation} \label{sec:impl}
% FPGA development at RTL level (Verilog / VHDL) usually requires a significant amount of human effort, limiting the applicability of FPGA-based accelerators.
% This is due to the intricacy of the algorithms, the necessity of manually tuning the synthesized design to fully exploit the FPGA's flexibility and implicit parallelism inherent in the algorithm, and the time-consuming verification process.
% High-level synthesis (HLS) is an automated process for synthesizing customized hardware from high-level languages including C/C++, which is advantageous and attractive in terms of improved productivity and rapid prototyping.
We implemented a scan matcher IP core that performs the aforementioned Greedy Endpoint Matching algorithm in parallel using Xilinx Vivado HLS v2019.2 toolchain.
We chose Pynq-Z2 development board as a target device (Table \ref{tbl:pynq-z2}), which is equipped with a programmable logic and a dual-core embedded processor, to demonstrate that the proposed core can be implemented in devices with severe resource constraints.
The clock frequency of the IP core is set to 100 MHz.

\begin{table}[h]
    \centering
    \caption{Specifications of Pynq-Z2 Board}
    \label{tbl:pynq-z2}
    \begin{tabular}{l|l} \hline \hline
        OS & Pynq Linux (based on Ubuntu 18.04) \\
        CPU & ARM Cortex-A9 @ 650MHz $\times$ 2 \\
        FPGA & Xilinx Zynq XC7Z020-1CLG400C (Artix-7) \\
        DRAM & 512MB (DDR3) \\ \hline \hline
    \end{tabular}
\end{table}

\begin{figure}[h]
    \centering
    \includegraphics[width=0.8\linewidth]{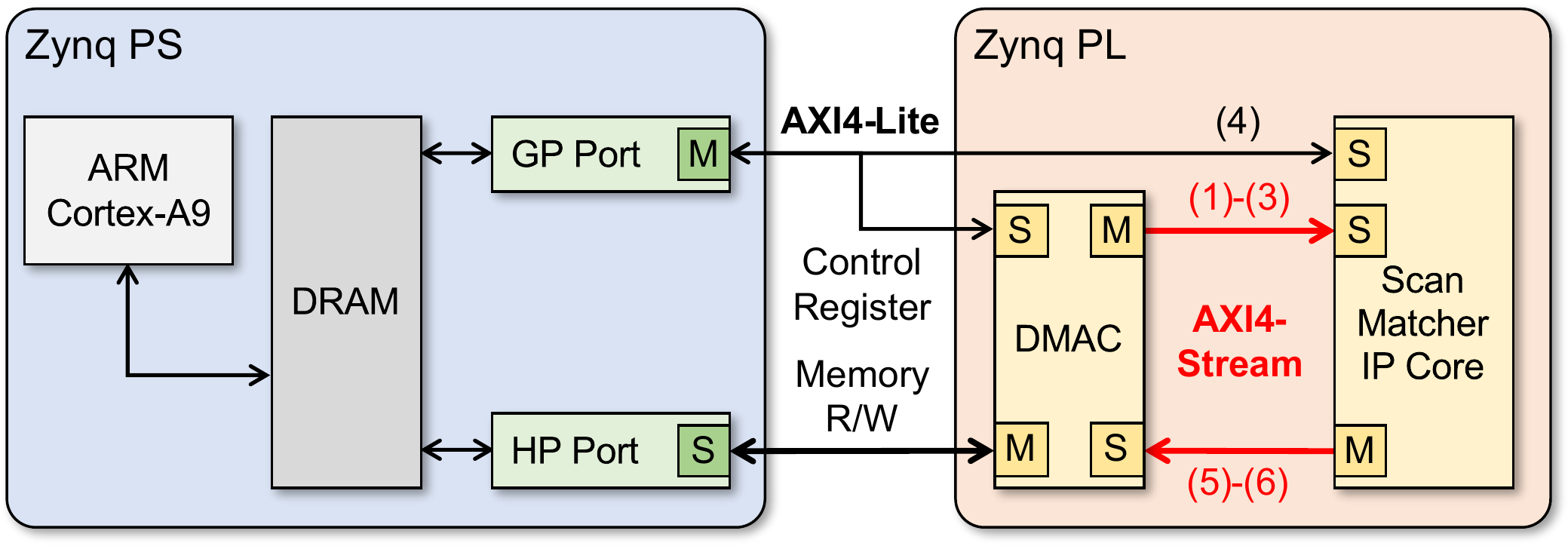}
    \caption{Board Level Implementation}
    \label{fig:board-level-impl}
\end{figure}

Figure \ref{fig:board-level-impl} depicts a brief overview of the board-level implementation.
The Zynq processing system (PS) executes our software implementation of GMapping algorithm (described in Section \ref{sec:pre}) except the scan matching part, which is offloaded to the programmable logic (PL) portion.
The PS passes the input data by communicating with the DMA controller to initiate the scan matcher IP core.
The DMA controller automatically creates fixed-sized AXI4-Stream packets containing the input data on the DRAM and delivers them to the IP core.
It also receives the AXI4-Stream packets returned from the IP core and writes the extracted result data to the specified address range of the DRAM.

The IP core takes the following inputs from the PS: (1) initial guess of the $N$ particle poses $\{ x_t'^{[k]} \}$, (2) $N$ local maps $\{ \widetilde{m}^{[k]} \}$, (3) the latest sensor measurements $z_t = \{ z_t^i \}$, and (4) additional parameters, where $N$ is a parallelization degree.
(4) includes the relative position of the local map $\widetilde{m}^{[k]}$ with respect to the entire map $m^{[k]}$.
The IP core then sends back (5) $N$ refined particle poses $\{ x_t^{[k]} \}$ and (6) final score values $\{ s(x_t^{[k]}, m^{[k]}, z_t) \}$ associated to $N$ particles to the PS; the latter can be used for weight computation.
To complete the scan matching process for all particles, the IP core should be repetitively invoked for $M / N$ times, where $M$ is the total number of particles used.
The input data that is shared among all particles (i.e. (3) and (4)) is transferred only once at the beginning of the scan matching phase.
The DMA controller makes use of a high-performance port (HP Port) on the board and also adopts AXI4-Stream protocol for high-speed transmission of most of the input (1)-(3) and output (5)-(6).
The other necessary parameters (4) are transferred via AXI4-Lite interface.
At the beginning of the software implementation, the bitstream (binary image) of the IP core design is dynamically loaded to the PL using Linux kernel FPGA manager.

\begin{figure}[h]
    \centering
    \includegraphics[width=0.8\linewidth]{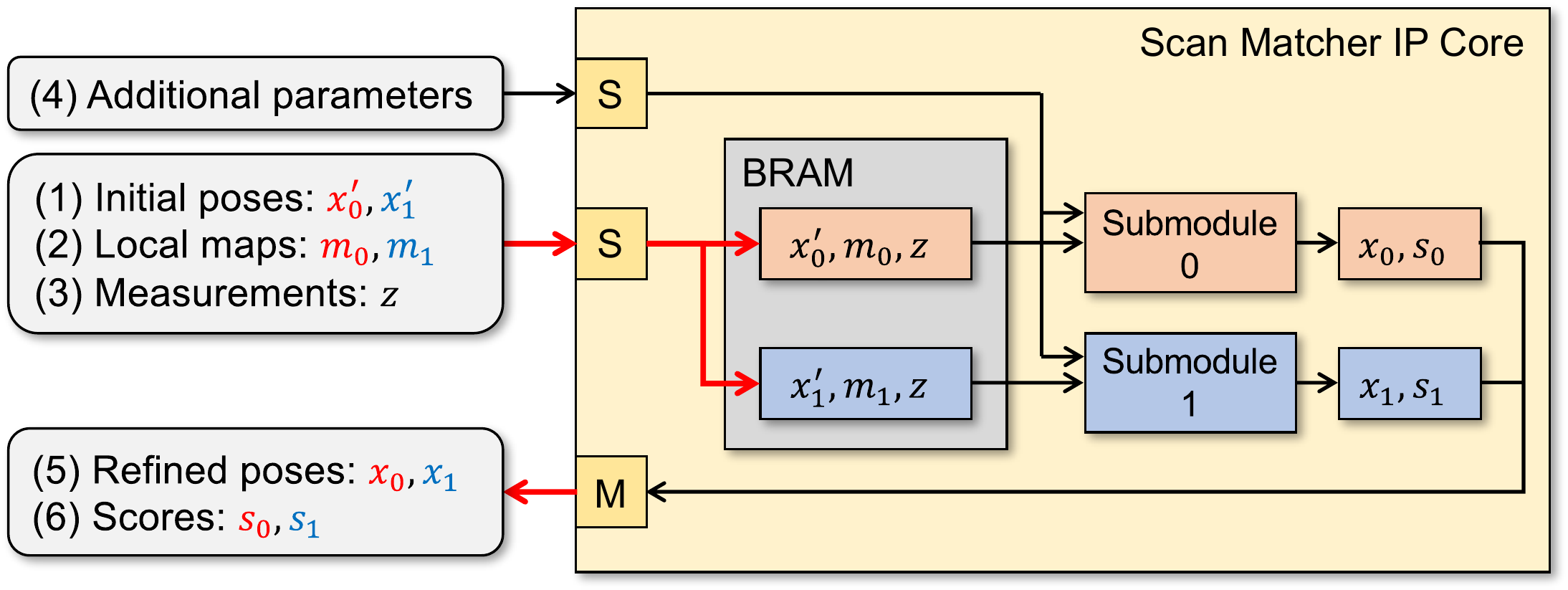}
    \caption{Design of Scan Matcher IP Core}
    \label{fig:scan-matcher-ip-core}
\end{figure}

Figure \ref{fig:scan-matcher-ip-core} illustrates the block diagram of the proposed scan matcher core.
The top module consists of two sub-modules, each of which computes the refined pose $x_t^{[k]}$ and the score value $s(x_t^{[k]}, m^{[k]}, z_t)$ for a single particle based on the Greedy Endpoint Matching algorithm, given the initial pose $x_t'^{[k]}$ and the local map $\widetilde{m}^{[k]}$.
As a result, the IP core performs the scan matching for two particles at the same time, resulting in a parallelization degree of $N = 2$.
Throughout our implementation, all the decimal numbers are represented by 32-bit fixed-point format with 16-bit signed integer and 16-bit fractional parts.
These bitwidths are determined to preserve the adequate precision for values such as the linear and angular component of particle poses; however, the search for the optimal fixed-point number expression depends on a given application (or a surrounding environment) and is beyond the scope of this paper.

It is worth mentioning that, in the software implementation of the scan matching, the particle pose is repeatedly updated until it satisfies the convergence condition (i.e. update step of the particle pose is below the preset threshold, or the number of iterations exceeds the maximum).
Conversely, in our IP core, the number of the optimization iterations is fixed (e.g. 25) in order to equalize the computational loads (latency) of all particles and realize the parallel execution.
It is one of the (4) additional parameters as noted above and thus can be set from the processing system before invoking the IP core.
We set this to 25 in all evaluations conducted in Section \ref{sec:eval}.
Accordingly, the IP core maintains constant latency cycles as long as the number of particles is kept.
Although this limitation typically causes the undesirable accuracy loss of the results, we observed that in most cases, the number of iterations is less than 25-30 and the average is around 10-15.
Also, we did not see a significant degradation in terms of accuracy as shown in Section \ref{sec:eval}.

% eval.tex

\section{Evaluations} \label{sec:eval}
In this section, the proposed scan matcher IP core is evaluated in terms of algorithm latency, accuracy, FPGA resource utilization, and power consumption in comparison with the software implementation.

\subsection{Experimental Setup} \label{sec:experimental-setup}
As a baseline, the entire GMapping algorithm is executed only with a CPU (ARM Cortex-A9 processor), which is denoted as \textbf{CPU$M$} (CPU, $M$ particles) in this experiment.
Then, the algorithm is executed with the CPU in cooperation with our IP core; that is, the CPU executes the software implementation of GMapping except the scan matching part, which is handled by our IP core.
We refer to this experimental setting as \textbf{FPGA$M$} (FPGA, $M$ particles).
The software is developed in C++ and compiled using GCC 7.3.0 with \texttt{-O3} compiler flag to fully optimize the executable code.

The subset of publicly available Radish dataset~\cite{RadishDataset}, namely Intel Research Lab (\textbf{Intel}, $28.5 \mathrm{m} \times 28.5 \mathrm{m}$), ACES Buliding (\textbf{ACES}, $56 \mathrm{m} \times 58 \mathrm{m}$), and MIT CSAIL Building (\textbf{MIT-CSAIL}, $61 \mathrm{m} \times 46.5 \mathrm{m}$) is used for the benchmarking purpose.
We chose these three datasets since they capture relatively small environments in which we expect our system to be run.
The ground truth information is unavailable in these datasets; they only contain the sequence of sensor observations and odometry robot poses, making quantitative comparisons difficult.
To measure the accuracy of output results (robot trajectories), we adopt the following performance metric proposed in \cite{Kuemmerle09}.
\begin{eqnarray}
    \varepsilon_{t - 1, t} &=& (x_t \ominus x_{t - 1}) \ominus \delta_{t - 1, t}^*
    \label{eq:metric-error} \\
    \varepsilon_\mathrm{trans} &=& \frac{1}{T} \sum_t
    || \mathrm{trans}(\varepsilon_{t - 1, t}) || \\
    \varepsilon_\mathrm{rot} &=& \frac{1}{T} \sum_t
    | \mathrm{rot}(\varepsilon_{t - 1, t}) | \\
    \sigma_\mathrm{trans}^2 &=& \frac{1}{T} \sum_t \left(
    || \mathrm{trans}(\varepsilon_{t - 1, t}) || - \varepsilon_\mathrm{trans} \right)^2 \\
    \sigma_\mathrm{rot}^2 &=& \frac{1}{T} \sum_t \left(
    | \mathrm{rot}(\varepsilon_{t - 1, t}) | - \varepsilon_\mathrm{rot} \right)^2,
\end{eqnarray}
where $\ominus$ denotes the inverse composition operator, i.e. $x \ominus y$ represents the relative transformation between two poses $x$ and $y$.
Two helper functions $\mathrm{trans}(x)$ and $\mathrm{rot}(x)$ split a given pose $x = [\xi_x, \xi_y, \xi_\theta]^\top$ into two translational components ($\xi_x, \xi_y$) and an angular component ($\xi_\theta$).
$|| \cdot ||$ is a norm function ($\sqrt{\xi_x^2 + \xi_y^2}$) and $| \cdot |$ is an absolute value ($|\xi_\theta|$).
The above metric computes the average and the standard deviation of discrepancies between two relative poses $x_t \ominus x_{t - 1}$ and $\delta_{t - 1, t}^*$; the former is the relative pose between temporally adjacent poses $x_t$ and $x_{t - 1}$, both of which are obtained from the trajectory result $x_{1 : T}$.
The latter is the ground truth relation extracted by manually matching the sensor observations (available at \cite{SLAMBenchmarking}).
We also use the above metric to evaluate the difference (closeness) between the trajectories obtained from CPU$M$ and FPGA$M$ to confirm that our scan matcher IP core achieves competitive accuracy compared to the software implementation.
We just substitute the $\delta_{t - 1, t}^*$ in Equation (\ref{eq:metric-error}) with the relative pose $\widehat{x}_t \ominus \widehat{x}_{t - 1}$, where $x_{1 : T}$ and $\widehat{x}_{1 : T}$ denote the trajectories from CPU$M$ and FPGA$M$, respectively.

\subsection{Algorithm Latency} \label{sec:eval-algorithm-latency}
\begin{figure}[h]
    \centering
    \includegraphics[width=0.7\linewidth]{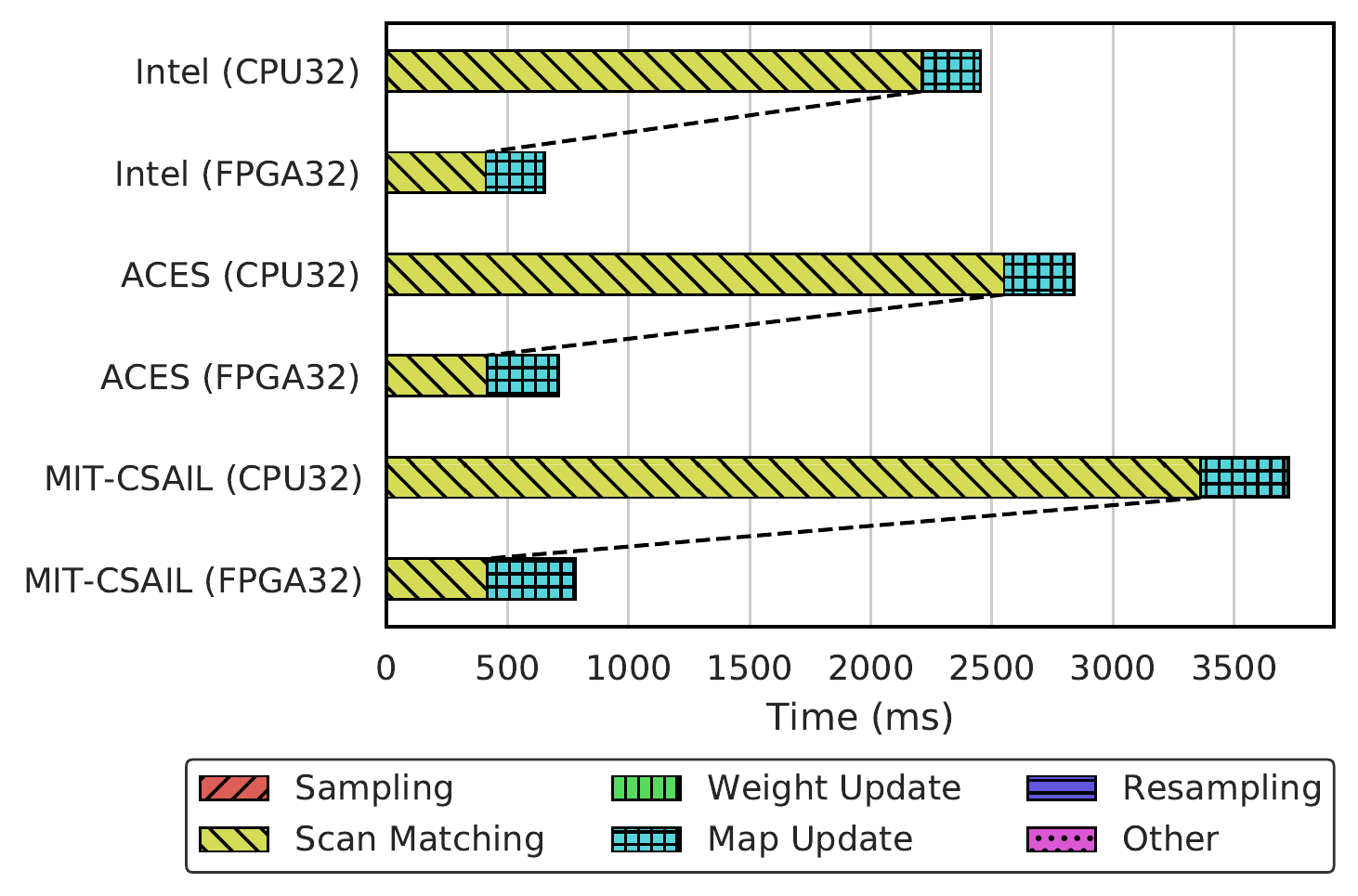}
    \caption{Comparison of Latency ($M = 32$)}
    \label{fig:breakdown-32}
\end{figure}

Figure \ref{fig:breakdown-32} shows the breakdown of the latency for a single iteration of the GMapping algorithm under two experimental configurations (CPU32 and FPGA32).
The results presented here are the average of 5 executions.
Note that the CPU-FPGA data transfer overhead is included in the scan matching latency for a fair comparison.
We observed that most of the execution time is dominated by scan matching and map update processes; other processes contribute a negligible amount to the latency.
The overall latency is effectively reduced up to $\times 4.77$ (MIT-CSAIL) as a result of offloading the costly scan matching computations to the FPGA.
For instance, in the Intel dataset, scan matching process accounts for 90.0 \% of the total runtime in CPU32, representing a major bottleneck, while it accounts for 62.8 \% in FPGA32.
Though we adopted the high-performance streaming protocol, the data transfer still accounts for a large proportion of the scan matching latency, which we attribute to the memory-mapped I/O used to access the DMA controller registers or to handle input/output data.
This indicates that if the data transfer overhead is minimized, the speedup ratio approaches to its theoretical maximum ($10.0 \times$ in the Intel case).

\begin{figure}[h]
    \centering
    \includegraphics[width=0.7\linewidth]{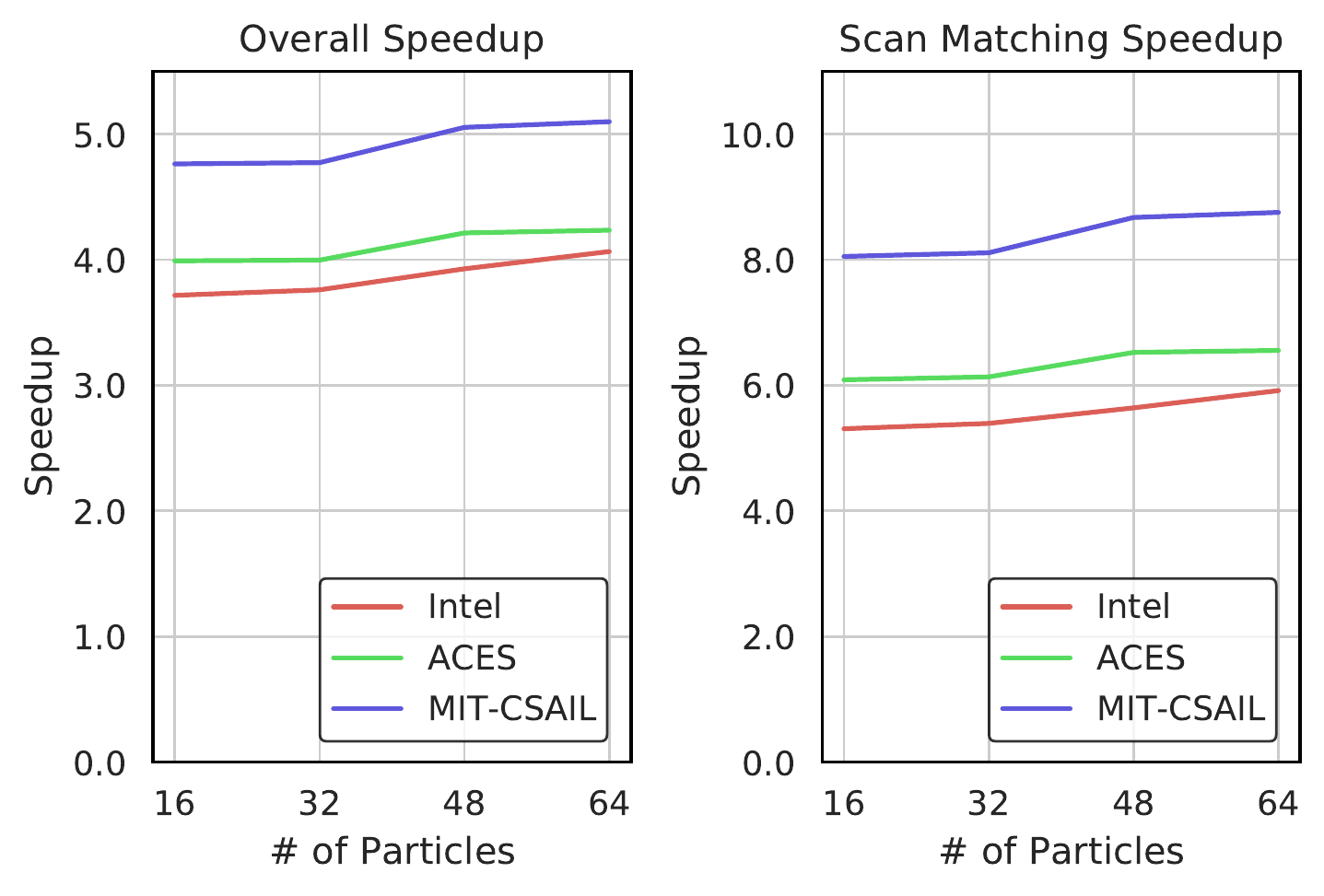}
    \caption{Relationship between Number of Particles and Speedup}
    \label{fig:speedup}
\end{figure}

The relationship between the number of particles $M$ and the speedup ratio is plotted in Figure \ref{fig:speedup}.
Our hardware implementation achieves the approximately constant speedup but with slight increase (6.09--6.56$\times$ for ACES, 5.31--5.92$\times$ for Intel, and 8.05--8.75$\times$ for MIT-CSAIL) under the varying number of particles, thus demonstrating the scalability of our proposed system.
The best speedup effect is obtained in the MIT-CSAIL dataset, in which the longest time is spent for scan matching computations among three datasets in the software implementation, while the latency of scan matching in our IP core remains constant regardless of the dataset used (see Section \ref{sec:impl}).
We observed the same behavior in the overall speedup (3.99--4.24$\times$ for ACES, 3.72--4.07$\times$ for Intel, and 4.76--5.10$\times$ for MIT-CSAIL).
Note that the slight increase of the speedup noticeable in three datasets comes from the slight performance degradation in the software implementation; we speculate that the consecutive accesses to the grid maps for a relatively large number of particles leads the increased cache miss rates in CPU.
In MIT-CSAIL dataset, the scan matching latency for a single particle is 104.18 ms and 113.06 ms when $M = 16$ and $M = 64$, respectively, which means the increase of latency by 8.5 \%.
% We expected that the latency of scan matching per one particle reduces as the number of particles increases.
% We transfer the input data that is shared among all particles (i.e. sensor measurements and algorithm parameters) only once at the beginning of the scan matching phase, meaning that the data transmission cost per one particle is alleviated when $M$ increases.
% However, we did not observe such an effect except the Intel dataset.
% For instance, the latency for a single particle is reduced from 13.02 ms ($M = 16$) to 12.76 ms ($M = 64$) in Intel dataset.

\subsection{Algorithm Accuracy} \label{sec:eval-accuracy}
The accuracy of the output trajectories is measured based on the metric proposed in \cite{Kuemmerle09}.
Table \ref{tbl:accuracy-comparison} compares the accuracy $\varepsilon_\mathrm{trans} \pm \sigma_\mathrm{trans}, \varepsilon_\mathrm{rot} \pm \sigma_\mathrm{rot}$ obtained from FPGA32 against CPU32 as a counterpart.
This result presents the favorable performance of the FPGA32 except for the translational error in ACES dataset, which is due to its environmental characteristics.
ACES dataset mainly consists of long straight corridors, which makes the results of the scan matching (i.e. refined poses) unreliable; that is, the positional uncertainty in the longitudinal direction of the corridor becomes inevitably large.
FPGA32 is more likely to suffer from the occurrence of the unreliable scan matching than CPU32, since it uses the fixed-point representation for decimal values in the scan matching process, which introduces the propagation and accumulation of rounding errors in addition to the quickly accumulating positional errors.
Table \ref{tbl:trajectory-comparison} shows the difference (closeness) between the trajectories obtained from CPU32 and FPGA32, which is computed by slightly modifying Equation (\ref{eq:metric-error}) as explained above.
Considering the map resolution ($\Delta = 0.05 \mathrm{m}$) and the angular resolution of the laser scanner ($0.5^\circ, 1.0^\circ$), it is obvious that the difference between two output trajectories is sufficiently small.
The translational difference does not surpass 0.1 m, which is equivalent to only two grid cells in a row.

\begin{table}[h]
    \centering
    \caption{Comparison of Accuracy ($M = 32$)}
    \label{tbl:accuracy-comparison}
    \begin{tabular}{c|cc} \hline \hline
        & CPU32 & FPGA32 \\ \hline
        ACES (m) & $0.0558 \pm 0.0649$ & $0.125 \pm 0.490$ \\
        ACES (rad) & $0.0851 \pm 0.319$ & $0.0852 \pm 0.319$  \\ \hline
        Intel (m) & $0.115 \pm 0.129$ & $0.117 \pm 0.130$ \\
        Intel (rad) & $0.0860 \pm 0.284$ & $0.0859 \pm 0.284$ \\ \hline
        MIT-CSAIL (m) & $0.0483 \pm 0.0764$ & $0.0505 \pm 0.0795$ \\
        MIT-CSAIL (rad) & $0.0970 \pm 0.387$ & $0.0984 \pm 0.387$ \\ \hline \hline
    \end{tabular}
\end{table}

\begin{table}[h]
    \centering
    \caption{Difference of Output Trajectories ($M = 32$)}
    \label{tbl:trajectory-comparison}
    \begin{tabular}{c|c} \hline \hline
        % & Average, Stddev & Median \\ \hline
        ACES (m) & $0.0712 \pm 0.127$ \\ % & $0.0352$
        ACES (rad) & $0.00765 \pm 0.00723$ \\ \hline % & $0.00549$
        Intel (m) & $0.0505 \pm 0.0705$ \\ % & $0.0316$
        Intel (rad) & $0.0134 \pm 0.0266$ \\ \hline % & $0.00716$
        MIT-CSAIL (m) & $0.0495 \pm 0.0641$ \\ % & $0.0324$
        MIT-CSAIL (rad) & $0.0106 \pm 0.0273$ \\ \hline \hline % & $0.00637$
    \end{tabular}
\end{table}

\begin{figure}[h]
    \centering
    \includegraphics[width=0.6\linewidth]{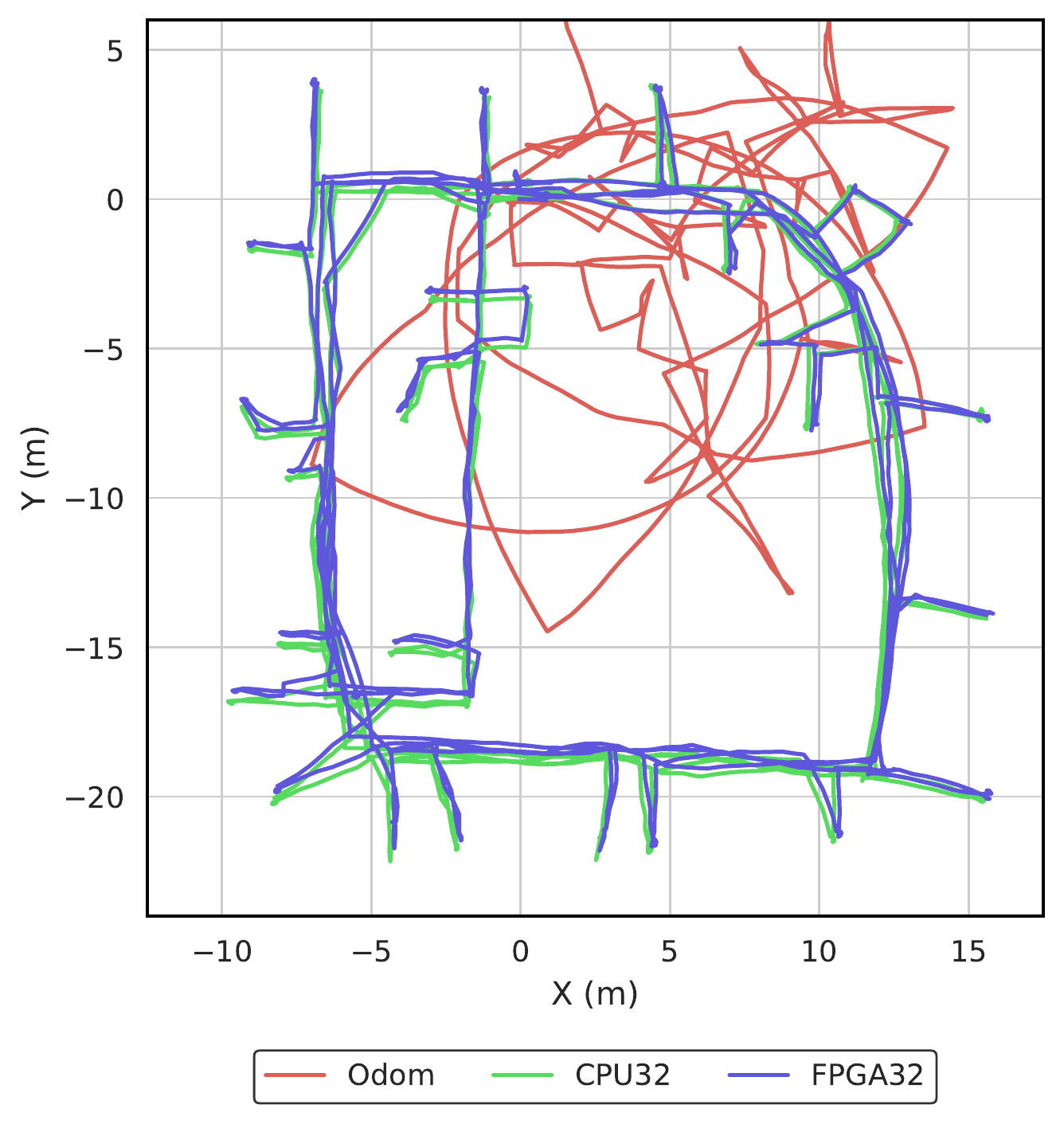}
    \caption{Trajectories Obtained from Intel Research Lab Dataset}
    \label{fig:trajectory-intel}
\end{figure}

Figure \ref{fig:trajectory-intel} shows the robot trajectories obtained from CPU32 and FPGA32.
The figure also shows the pure odometry trajectory, denoted as \textbf{Odom}.
The considerable overlap between CPU32 and FPGA32 implies that the accuracy is not severely affected by introducing local maps as a part of map compression technique (Section \ref{sec:map-compression}).
The scan points (obstacles) outside the local map are ignored in the score evaluations (Equation (\ref{eq:scan-matching-score})), which causes erroneous scan matching results especially when local maps are too small.
In FPGA32, the computation based on fixed-point expressions introduces rounding errors, which would serve as a primary source of precision loss.
FPGA32 is also affected by the limitation of the number of algorithm iterations (Section \ref{sec:impl}), by which the robot pose is not fully optimized and hence the cumulative error grows rapidly.
Contrary to these concerns, FPGA32 still generates the topologically correct map and the underlying geometric relationship is maintained.
In addition, the distortion and imprecision caused by these factors seem subtle, which is the satisfying outcome.

\subsection{FPGA Resource Utilization} \label{sec:eval-resource}
Table \ref{tbl:resource-consumption} shows the FPGA resource utilization of our implementation, designed for Xilinx Zynq XC7Z020-1CLG400C assuming 100 MHz operating frequency.
On-chip BRAMs are mostly consumed for the storage of local maps to execute the scan matching for multiple particles simultaneously, which implies that the BRAM consumption increases almost linearly proportional to the degree of parallelization.
In our current design, the scan matching is parallelized for two particles, and the BRAM usage is still less than 50 \% due to the map compression technique as described in Section \ref{sec:map-compression}.
Especially, the extreme quantization of the occupancy value contributes to the resource reduction.
The design uses certain amount of the LUT slices since mathematical operations (coordinate transformations) are frequently performed on the core.
Though the achievable speedup is constrained by the total amount of BRAM and LUT resources present on a device, results in Table \ref{tbl:resource-consumption} suggest that other parts of the algorithm (i.e. importance weight calculation and initial pose guess) can be mapped onto the hardware.
There is also enough room to increase the parallelization degree (e.g. 4) to achieve the further performance improvement.

\begin{table}[h]
    \centering
    \caption{FPGA Resource Utilization of Scan Matcher IP Core (Post Place-and-Route)}
    \label{tbl:resource-consumption}
    \begin{tabular}{c|rrrr} \hline \hline
        & BRAM & DSP & FF & LUT \\ \hline
        Total & 61 & 32 & 18,887 & 23,254 \\
        Available & 140 & 220 & 106,400 & 53,200 \\
        Utilization (\%) & 43.6 & 14.6 & 17.8 & 43.7 \\ \hline \hline
    \end{tabular}
\end{table}

\subsection{Power Consumption} \label{sec:eval-power}
Our board-level implementation (FPGA32) consumed 2.9 W of power, which is as same as the software-only implementation (CPU32).
We used an ordinary watt-hour meter to measure the entire power consumption of the Pynq-Z2 board.
We emphasize that FPGA32 outperforms CPU32 in terms of the total execution time ($3.76\times$ shorter) when using Intel dataset as shown in Figure \ref{fig:speedup}.

% supposed to
% indistinguishable
% generalization
% elucidate
% superior

% conc.tex

\section{Summary} \label{sec:conc}
The hardware optimization of SLAM methods is of crucial importance for deploying SLAM applications to autonomous mobile robots with severe limitations in power delivery and available resources.
In this work, we proposed a lightweight FPGA-based design dedicated to accelerating the scan matching process in the 2D LiDAR SLAM method called GMapping by exploiting the parallel structure inherent in the algorithm.
The resource usage and the overhead associated with the data transfers are effectively reduced by applying the map compression technique, which is the combination of map binarization and introduction of local maps.
The map data is stored with the acceptable level of redundancy to enable the efficient data accesses thereby minimizing the latency.
Also, the precomputed lookup table is employed to eliminate the expensive mathematical computations.
% The precomputed lookup table is employed to enable the concise evaluation of the observation likelihood thereby improving the algorithm throughput.
% The expensive mathematical computation is eliminated in exchange for an affordable increase in resource consumption.
Experiments based on benchmark datasets demonstrated that our hardware scan matcher avoids the loss of accuracy and offers satisfactory throughput to that of the software implementation.
The proposed core achieved 5.31--8.75$\times$ scan matching speedup and 3.72--5.10$\times$ overall speedup.
As far as we know, this is the first work that focuses on the hardware acceleration of the grid-based RBPF-SLAM.
% As far as we know, this is the first work that focuses on the hardware acceleration of the grid-based RBPF-SLAM, which is an essential functionality to realize autonomous robots.

% However, there is still much room for further improvement.
% The implementation of the real hardware devices is mandatory to measure the runtime power consumption and to prove the practicality of our approach since the performance evaluation of the hardware module is currently conducted based on the software emulation.
% In this work, the likelihood field model-based algorithm called Greedy Endpoint Matching is implemented on the FPGA, which is susceptible to local optima.
% There exists a variety of scan matching methods proposed in the literature, therefore other well-known variants such as Iterative Closest Points (ICP) \cite{Besl92}, Correlative Scan Matching \cite{Olson09} and map-to-map based matching are worth investigating for the hardware implementation.
% It is possible to conduct a comprehensive comparison between these techniques in terms of accuracy, resource utilization, and energy consumption.
% Also, techniques for sharing maps between multiple particles \cite{Schroeter07}\cite{Jo18} can be employed to further reduce the memory consumption.

% Bibliography
\renewcommand{\baselinestretch}{1.0}
\bibliographystyle{unsrt}
% \bibliography{refer}

\begin{thebibliography}{10}

\bibitem{Dissanayake01}
M.W.M.G. Dissanayake, P.~Newman, S.~Clark, H.F. Durrant-Whyte, and M.~Csorba.
\newblock {A Solution to the Simultaneous Localisation and Map Building (SLAM)
  Problem}.
\newblock {\em IEEE Transactions on Robotics and Automation}, 17(3):229--241,
  June 2001.

\bibitem{Montemerlo02}
Michael Montemerlo, Sebastian Thrun, Daphne Koller, and Ben Wegbreit.
\newblock {FastSLAM 1.0: A Factored Solution to the Simultaneous Localization
  and Mapping Problem}.
\newblock In {\em Proceedings of the AAAI National Conference on Artificial
  Intelligence}, 2002.

\bibitem{Montemerlo03}
Michael Montemerlo, Sebastian Thrun, Daphne Koller, and Ben Wegbreit.
\newblock {FastSLAM 2.0: An Improved Particle Filtering Algorithm for
  Simultaneous Localization and Mapping that Provably Converges}.
\newblock In {\em Proceedings of the International Joint Conference on
  Artificial Intelligence (IJCAI)}, pages 1151--1156, June 2003.

\bibitem{Grisetti07B}
Giorgio Grisetti, Cyrill Stachniss, and Wolfram Burgard.
\newblock {Improved Techniques for Grid Mapping with Rao-Blackwellized Particle
  Filters}.
\newblock {\em IEEE Transactions on Robotics}, 23(1):32--46, March 2007.

\bibitem{Santos13}
João~Machado Santos, David Portugal, and Rui~P. Rocha.
\newblock {An evaluation of 2D SLAM techniques available in Robot Operating
  System}.
\newblock In {\em Proceedings of the IEEE International Symposium on Safety,
  Security, and Rescue Robotics (SSRR)}, pages 1--6, October 2013.

\bibitem{Hess16}
Wolfgang Hess, Damon Kohler, Holger Rapp, and Daniel Andor.
\newblock {Real-Time Loop Closure in 2D LIDAR SLAM}.
\newblock In {\em Proceedings of the IEEE International Conference on Robotics
  and Automation (ICRA)}, pages 1271--1278, 2016.

\bibitem{Nardi15}
Luigi Nardi, Bruno Bodin, M.~Zeeshan Zia, John Mawer, Andy Nisbet, Paul H.~J.
  Kelly, Andrew~J. Davison, Mikel Luj{\'a}n, Michael F.~P. O'Boyle, Graham
  Riley, Nigel Topham, and Steve Furber.
\newblock {Introducing SLAMBench, a performance and accuracy benchmarking
  methodology for SLAM}.
\newblock In {\em Proceedings of the IEEE International Conference on Robotics
  and Automation (ICRA)}, pages 5783--5790, May 2015.

\bibitem{Gautier19}
Quentin Gautier, Alric Althoff, and Ryan Kastner.
\newblock {FPGA Architectures for Real-time Dense SLAM}.
\newblock In {\em Proceedings of the IEEE International Conference on
  Application-specific Systems, Architectures and Processors (ASAP)}, pages
  83--90, July 2019.

\bibitem{Thrun05}
Sebastian Thrun, Wolfram Burgard, and Dieter Fox.
\newblock {\em {Probabilistic Robotics}}.
\newblock MIT Press, 2005.

\bibitem{Grisetti07A}
Giorgio Grisetti, Gian~Diego Tipaldi, Cyrill Stachniss, Wolfram Burgard, and
  Daniele Nardi.
\newblock {Fast and Accurate SLAM with Rao-Blackwellized Particle Filters}.
\newblock {\em Robotics and Autonomous Systems}, 55(1):30--38, January 2007.

\bibitem{Doucet00}
Arnaud Doucet, Nando de~Freitas, Kevin~P. Murphy, and Stuart~J. Russell.
\newblock {Rao-Blackwellised Particle Filtering for Dynamic Bayesian Networks}.
\newblock In {\em Proceedings of the Conference on Uncertainty in Artificial
  Intelligence (UAI)}, pages 176--183, 2000.

\bibitem{Kuemmerle09}
Rainer Kuemmerle, Bastian Steder, Christian Dornhege, Michael Ruhnke, Giorgio
  Grisetti, Cyrill Stachniss, and Alexander Kleiner.
\newblock {On measuring the accuracy of SLAM algorithms}.
\newblock {\em Autonomous Robots}, 27(4):387--407, November 2009.

\bibitem{Par11}
Kerem Par and O{\v{g}}uz Tosun.
\newblock {Parallelization of particle filter based localization and map
  matching algorithms on multicore/manycore architectures}.
\newblock In {\em Proceedings of the IEEE Intelligent Vehicles Symposium (IV)},
  pages 820--826, June 2011.

\bibitem{Llofriu13}
Martin Llofriu, Federico Andrade, Facundo Benavides, Alfredo Weitzenfeld, and
  Gonzalo Tejera.
\newblock {An embedded particle filter SLAM implementation using an affordable
  platform}.
\newblock In {\em Proceedings of the International Conference on Advanced
  Robotics (ICAR)}, pages 1--6, November 2013.

\bibitem{Gouveia15}
David Portugal, Bruno~D. Gouveia, and Lino Marques.
\newblock {A Distributed and Multithreaded SLAM Architecture for Robotic
  Clusters and Wireless Sensor Networks}.
\newblock {\em Studies in Computational Intelligence}, pages 121--141, May
  2015.

\bibitem{Abouzahir16}
Mohamed Abouzahir, Abdelhafid Elouardi, Samir Bouaziz, Rachid Latif, and
  Abdelouahed Tajer.
\newblock {Large-scale monocular FastSLAM2.0 acceleration on an embedded
  heterogeneous architecture}.
\newblock {\em EURASIP Journal on Advances in Signal Processing}, pages 1--20,
  2016.

\bibitem{Sileshi16}
B.G. Sileshi, J.~Oliver, R.~Toledo, J.~Gonçalves, and P.~Costa.
\newblock {On the behaviour of low cost laser scanners in HW/SW particle filter
  SLAM applications}.
\newblock {\em Robotics and Autonomous Systems}, pages 11--23, 2016.

\bibitem{Abouzahir18}
Mohamed Abouzahir, Abdelhafid Elouardi, Rachid Latif, and Samir Bouaziz.
\newblock {Embedding SLAM algorithms: Has it come of age?}
\newblock {\em Robotics and Autonomous Systems}, pages 14--26, 2018.

\bibitem{Wurm10}
Kai~M. Wurm, Cyrill Stachniss, and Giorgio Grisetti.
\newblock {Bridging the gap between feature- and grid-based SLAM}.
\newblock 58(2):140--148, February 2010.

\bibitem{Schroeter07}
Christof Schr{\"o}ter, Hans-Joachim B{\"o}hme, and Horst-Michael Gross.
\newblock {Memory-Efficient Gridmaps in Rao-Blackwellized Particle Filters for
  SLAM using Sonar Range Sensors}.
\newblock In {\em Proceedings of the European Conference on Mobile Robots
  (EMCR)}, September 2007.

\bibitem{Jo18}
HyungGi Jo, Hae~Min Cho, Sungjin Jo, and Euntai Kim.
\newblock {Efficient Grid-Based Rao-Blackwellized Particle Filter SLAM With
  Interparticle Map Sharing}.
\newblock {\em IEEE/ASME Transactions on Mechatronics}, 23(2):714--724, April
  2018.

\bibitem{Gouveia14}
Bruno~D. Gouveia, David Portugal, and Lino Marques.
\newblock {Speeding Up Rao-Blackwellized Particle Filter SLAM with a
  Multithreaded Architecture}.
\newblock In {\em Proceedings of the IEEE/RSJ International Conference on
  Intelligent Robots and Systems (IROS)}, pages 1583--1588, 2014.

\bibitem{QiuchengLi18}
Qiucheng Li, Thomas Rauschenbach, Andreas Wenzel, and Fabian Mueller.
\newblock {EMB-SLAM: An Embedded Efficient Implementation of Rao-Blackwellized
  Particle Filter Based SLAM}.
\newblock In {\em Proceedings of the International Conference on Control,
  Robotics and Cybernetics (CRC)}, pages 88--93, September 2018.

\bibitem{Olson15}
Edwin Olson.
\newblock {M3RSM: Many-to-Many Multi-Resolution Scan Matching}.
\newblock In {\em Proceedings of the IEEE International Conference on Robotics
  and Automation (ICRA)}, pages 5815--5821, 2015.

\bibitem{OpenSLAM}
Cyrill Stachniss, Udo Frese, and Giorgio Grisetti.
\newblock {OpenSLAM.org}.
\newblock \url{https://openslam-org.github.io/}, 2007.

\bibitem{Montemerlo03B}
Michael Montemerlo, Nicholas Roy, and Sebastian Thrun.
\newblock {Perspectives on Standardization in Mobile Robot Programming : The
  Carnegie Mellon Navigation (CARMEN) Toolkit}.
\newblock In {\em Proceedings of the IEEE/RSJ International Conference on
  Intelligent Robots and Systems (IROS)}, pages 2436--2441, 2003.

\bibitem{Olson09}
Edwin~B. Olson.
\newblock {Real-Time Correlative Scan Matching}.
\newblock In {\em Proceedings of the IEEE International Conference on Robotics
  and Automation (ICRA)}, pages 4387--4393, 2009.

\bibitem{RadishDataset}
Andrew Howard and Nicholas Roy.
\newblock {The Robotics Data Set Repository (Radish)}.
\newblock \url{http://radish.sourceforge.net/}, 2003.

\bibitem{SLAMBenchmarking}
Alexander Kleiner, Bastian Steder, Christian Dornhege, Cyrill Stachniss,
  Giorgio Grisetti, Michael Ruhnke, Rainer K{\"u}mmerle, and Wolfram Burgard.
\newblock {SLAM Benchmarking Home}.
\newblock \url{http://ais.informatik.uni-freiburg.de/slamevaluation/}, 2009.

\end{thebibliography}

\end{document}